\documentclass{article}[titlepage]

\usepackage{orcidlink,lmodern}
\usepackage[onehalfspacing]{setspace}
\usepackage{fullpage}
\usepackage{amsmath}
\usepackage{amssymb}
\usepackage{booktabs}
\usepackage{tabularx}
\usepackage{makecell}
\usepackage{lipsum}
\usepackage{natbib}
\usepackage{framed}
\usepackage{algorithm}
\usepackage{algpseudocode}
\usepackage{enumitem}

\newcommand{\fct}[1]{#1()}
\newcommand{\pkg}[1]{\textbf{#1}}
\newcommand{\code}[1]{`#1'}
\newcommand{\proglang}[1]{\textbf{#1}}
\graphicspath{{./fig/}}
\setlist[enumerate]{leftmargin=*, topsep=0pt, itemsep=0pt}
\setlist[itemize]{leftmargin=*, topsep=0pt, itemsep=0pt}

\author{Tianchen Xu~\orcidlink{0000-0002-0102-7630}\\Bristol Myers Squibb
\and Wen Zhang \\ Amgen Inc. \and Rachael Wen\\Bristol Myers Squibb }

\title{\pkg{PWEXP}: An \proglang{R} Package Using Piecewise Exponential Model for Study Design and Event/Timeline Prediction}

\begin{document}

\maketitle

\begin{abstract}
Parametric assumptions such as exponential distribution are commonly used in clinical trial design and analysis. However, violation of distribution assumptions can introduce biases in sample size and power calculations. Piecewise exponential (PWE) hazard model partitions the hazard function into segments each with constant hazards and is easy for interpretation and computation. Due to its piecewise property, PWE can fit a wide range of survival curves and accurately predict the future number of events and analysis time in event-driven clinical trials, thus enabling more flexible and reliable study designs. Compared with other existing approaches, the PWE model provides a superior balance of flexibility and robustness in model fitting and prediction.

The proposed \pkg{PWEXP} package is designed for estimating and predicting PWE hazard models for right-censored data. 
By utilizing well-established criteria such as AIC, BIC, and cross-validation log-likelihood, the \pkg{PWEXP} package chooses the optimal number of change-points and determines the optimal position of change-points. With its particular goodness-of-fit, the \pkg{PWEXP} provides accurate and robust hazard estimation, which can be used for reliable power calculation at study design and timeline prediction at study conduct. 
The package also offers visualization functions to facilitate the interpretation of survival curve fitting results.

\end{abstract}



\section[Introduction: Count data regression in R]{Introduction} \label{sec:intro}
Survival analysis is a statistical method used to analyze time-to-event data, such as the time from a subject's randomization until he/she experiences an event (e.g., disease progression or death). One of the most commonly parametric models in survival analysis is the exponential hazard model, which assumes a constant hazard rate over time. However, in reality, the hazard rate is barely constant over time. Abrupt or gradual changes in the hazard rates  can be observed due to environmental factors, specific maintenance activity, and so on. The change of hazard rates over time could happen multiple times \citep{qian2013multiple}. A more flexible model is needed and the piecewise exponential hazard model is a popular alternative.
The piecewise exponential (PWE) model partitions the hazard over time into several time intervals, assuming the hazard rates are constant within each time interval, but can vary between intervals. 
By selecting appropriate change-points to create closely-spaced boundaries  and dividing the time axis into smaller intervals where the hazard changes rapidly and wider intervals where the hazard changes slowly, the hazard rate within each time interval can be estimated and their estimations can be used to approximate the Kaplan-Meier (KM) survival curves. Such approximation capability is due to its inherent flexibility which allows it to effectively capture a wide range of hazard rate patterns \citep{matthews1982testing}.

Estimated piecewise hazard rates from from historical data can provide assumptions for clinical trial design, including sample size power calculation, interim analysis (IA) plan, trial duration estimation, and resource allocation, etc. Commonly used packages for clinical trial design such as \pkg{rpact} \citep{rpact}, \pkg{gsDesign} \citep{geDesign} can utilize piecewise exponential models, which will be illustrated in later sections. 

\begin{table}[t!]
\centering
\newcolumntype{Y}{>{\centering\arraybackslash}X}
\begin{tabularx}{1.05\textwidth}{lYYYY>{\centering\arraybackslash}p{61pt}}
\toprule
& \multicolumn{5}{c}{\proglang{R} Package} \\
\cmidrule{2-6}
 & \pkg{PWEXP} & \pkg{pch}   & \pkg{muhaz} & \pkg{eventTrack} & \makecell{\pkg{Piecewise-}\\ \pkg{Changepoint}} \\
\midrule
Method & MLE/OLS & MLE   & MLE   & PML   & Bayesian \\
Model estimation with&       &       &       &       &  \\
\hspace{1em}Given change-points & \checkmark & \checkmark & $\times$ & $\times$ & $\times$ \\
\hspace{1em}Unknown change-points & \checkmark & $\times^{(a)}$ & $\times^{(b)}$ & $\checkmark^{(c)}$  & \checkmark \\
\hspace{1em}Partial known change-points & \checkmark & $\times$ & $\times$ & $\times$ & $\times$ \\
\hspace{1em}Tail robustness adjustment & $\checkmark^{(d)}$ & $\times$ & $\times$ & $\times$ & $\times$ \\
Inference &       &       &       &       &  \\
\hspace{1em}CI/Asymptotic variance & \checkmark & \checkmark & $\times$ & \checkmark &  $\checkmark^{(e)}$   \\
\hspace{1em}Change-point test & $\times$ & $\times$ & $\times$ & \checkmark & $\times$ \\
Model summary &       &       &       &       &  \\
\hspace{1em}AIC, BIC & \checkmark & \checkmark & $\times$ & $\times$ & $\times$ \\
\hspace{1em}PML, WAIC$^{(f)}$ & $\times$ & $\times$ & $\times$ & $\times$ & \checkmark \\
\hspace{1em}Likelihood & \checkmark & \checkmark & $\times$ & \checkmark & \checkmark \\
\hspace{1em}CV likelihood & \checkmark & $\times$ & $\times$ & $\times$ & $\times$ \\
\bottomrule
\end{tabularx}%
\caption{Comparison between the \pkg{PWEXP} package and other alternatives for PWE model estimation. The symbol \checkmark indicates the feature is available, and $\times$ indicates it is not available. Note: (a) The \pkg{pch} package uses empirical quantiles as change-points; (b) The \pkg{muhaz} package uses even length intervals for change-points; (c) The \pkg{eventTrack} uses general-purpose optimizer ``Nelder-Mead Simplex'' method to estimate change-points; (d) The \pkg{PWEXP} provides several approaches to avoid an unstable tail (See Section~\ref{sec:tail}); (e) Credible interval; (f) PML: Pseudo-Marginal Likelihood, WAIC: Widely Applicable Information Criterion.}\label{tab:pwe} 
\end{table}

Over the past decades, significant progress has been made in developing mature theories regarding the asymptotic properties of estimators in PWE models. 
\citet{friedman1982piecewise, yao1986maximum} demonstrated that the maximum likelihood (ML) change-point estimator is consistent under certain conditions. 
\citet{henderson1990problem} derived exact critical
values for likelihood ratio test for a single change-point and the corresponding piecewise exponential hazard rates. \citet{goodman2011detecting} proposed a Wald-type test statistic to test the existence of multiple change-points in a PWE model. 

\begin{table}[!t]
\centering
\newcolumntype{Y}{>{\centering\arraybackslash}X}
\begin{tabularx}{\textwidth}{lYcYY}
\toprule
& \multicolumn{4}{c}{\proglang{R} Package} \\
\cmidrule{2-5}
  & \pkg{PWEXP} & \pkg{eventTrack} & \pkg{gestate} & \pkg{eventPred} \\
\midrule
Sub-model &       &       &       &  \\
\hspace{1em}Accural model & uniform/user defined & user defined & user defined & (piecewise) Poisson/time-decay/B-spline \\
\hspace{1em}Event model & exp/pwexp & KM+exp tail & exp/Weibull/log-normal & various$^{(a)}$ \\
\hspace{1em}Drop-out model & exp/pwexp & $\times$ & exp/Weibull/log-normal/other$^{(b)}$ & various$^{(a)}$ \\
Prediction &       &       &       &  \\
\hspace{1em}Event prediction  & \checkmark & \checkmark & \checkmark & \checkmark \\
\hspace{1em}Timeline prediction  & \checkmark & $\times$ & $\times$ & \checkmark \\
\hspace{1em}Confidence interval    & \checkmark & $\times$ & \checkmark&  $\checkmark^{(c)}$ \\
\bottomrule
\end{tabularx}
\caption{Comparison between the \pkg{PWEXP} package and other alternatives for event/timeline prediction. The symbol \checkmark indicates the feature is available, and $\times$ indicates it is not available. Note: (a) The \pkg{eventPred} includes these event/drop-out models: exponential, Weibull, log-logistic, log-normal, piecewise exponential (with known change-points), model averaging, spline; (b) These drop-out distributions can be used in the \pkg{gestate} package as known models (since the package cannot estimate them from data): generalized Gamma, Gompertz, log-logistic, mixture exponential, mixture Weibull, piecewise exponential; (c) The \pkg{eventPred} package does not account for the uncertainty from model estimation.}\label{tab:prediction} 
\end{table}

However, despite the significant progress in PWE model inference, there are still limited approaches available to accurately estimate the change-points in real datasets. 
For a PWE model with one single change-point, it is recommended to employ grid search or solve the score equations using numerical algorithms like Newton-Raphson \citep{dupuy2006estimation, li2013estimation}.
For a PWE model with multiple change-points , \citet{goodman2011detecting} maximum (profile) likelihood estimation with the Nelder-Mead Simplex optimization algorithm is applicable. However, since the log-likelihood function is highly  non-smooth, the recommended method heavily depends on the initial values and is often very slow and unstable in practice. 
There are also several Bayesian methods available, including stochastic approximation Monte Carlo algorithm\citep{kim2020bayesian}, reversible jump MCMC method and some variants \citep{chapple2020novel, cooney2021change}.

The proposed \pkg{PWEXP} offers several ways to estimate PWE models ranging from a single change-point to multiple change-points, with the number and positions of change-points know or not known. Compared with available packages on CRAN for PWE model estimation \citep{pch, muhaz, eventTrack, PiecewiseChangepoint}, \pkg{PWEXP} stands out among these packages by goodness-of-fit, robust estimation of varied survival curves, and offering a comprehensive toolset (Table~\ref{tab:pwe}).

Another primary focus of \pkg{PWEXP} is the event/timeline prediction in clinical trials.  In event-driven clinical trials, accurately predicting future analysis time or events by using interim data is crucial for resource allocation, informed decision-making, and effective program development. 
Following the classic event prediction framework \citep{bagiella2001predicting, ying2008weibull}, at the current calendar time, $t_0$, we are interested in predicting the conditional expectation of the number of events $ED(t_0,t')$ that will have been observed by time $t'$ ($t'>t_0$): 
$ED(t_0,t)=D(t_0)+Q(t_0,t')+R(t_0,t')$, where $D(t_0)$ is the total number of subjects with events by time $t_0$; $Q(t_0, t')$ is the expected number of subjects who are in the study with no events by time $t_0$ and will have events by time $t'$; and $R(t_0,t')$ is the expected number of subjects who will enroll and have events between times $t_0$ and $t'$. 
If the trial has finished enrollment by $t_0$, then $R(t_0,t')=0$. 
Among the three components, $D(t_0)$ is always known; $Q(t_0, t')$ and $R(t_0,t')$ requires modeling on the distribution of events and loss to follow-up (drop-out) (e.g., using PWE models). For $R(t_0,t')$, an additional accrual prediction model is also required.
Table~\ref{tab:prediction} lists existing \proglang{R} packages for event/timeline prediction. \pkg{PWEXP} package offers both event and timeline prediction along with confidence intervals.
The \pkg{eventPred} can also take PWE model as event and drop-out model, but it cannot estimate a model without known change-points.

In the remainder of this article, Section~\ref{sec:background} introduces the theories and estimation approaches of PWE model, Section~\ref{sec:prediction} focuses on the event prediction framework, Section~\ref{sec:package} demonstrates the \pkg{PWEXP} package, followed with Section~\ref{sec:simulation} presents a general analysis workflow of event prediction by using simulated data,  Section~\ref{sec:design} illustrates how to use \pkg{PWEXP} for study design with a real data example, and the last section is for summary and conclusion. 

\section{Piecewise Exponential (PWE) Model} \label{sec:background}
\subsection{PWE Distribution and Survival Model}\label{sec:loglike}
This section begins with a brief introduction to the Piecewise Exponential (PWE) distribution, followed by an overview of the PWE model for right censoring data.

When the hazard rates of a random variable $T$ is piecewise constant with a total of $r$ change-points $d_k$ ($1\le k\le r$):  
\begin{align*}
  h(t)&=\begin{cases}
    \lambda_1, & t<d_1\\
    \lambda_2, & d_1\le t<d_2\\
    \cdots\\
    \lambda_{r+1}, & t\ge d_r
\end{cases}, 
\end{align*}
then $T$ follows a piecewise exponential (PWE) distribution. 

The density function and survival function ($1-$cumulative density function) of $T$ are given by:
\begin{align*}
  f(t)=h(t)e^{-H(t)}&=\begin{cases}
    \lambda_1 e^{-\lambda_1 t}, & t<d_1\\
    \lambda_2 e^{(\lambda_2-\lambda_1)d_1-\lambda_2t} , & d_1\le t<d_2\\
    \cdots\\
    \lambda_{r+1}e^{\left[\sum_{i=1}^r(\lambda_{i+1}-\lambda_{i})d_i\right]-\lambda_{r+1}t}, & t\ge d_r
\end{cases}\\
S(t)=e^{-H(t)}&=\begin{cases}
    e^{-\lambda_1 t}, & t<d_1\\
    e^{(\lambda_2-\lambda_1)d_1-\lambda_2t} , & d_1\le t<d_2\\
    \cdots\\
    e^{\left[\sum_{i=1}^r(\lambda_{i+1}-\lambda_{i})d_i\right]-\lambda_{r+1}t}, & t\ge d_r.
\end{cases}
\end{align*}
The \fct{dpwexp} and \fct{ppwexp} functions in the \pkg{PWEXP} package can be used to calculate these values.

A PWE random variable $T$ can be generated with inverse transform method, which relies on the quantile function $Q(p)$:
\begin{align*}
  Q(p)&=\begin{cases}
      \frac{-\log(1-p)}{\lambda_1} & p< 1-e^{-\lambda d_1}\\
      \frac{(\lambda_2-\lambda_1)d_1-\log(1-p)}{\lambda_2} & 1-e^{-\lambda d_1}\le p< 1-e^{(\lambda_2-
      \lambda_1)d_1-\lambda_2 d_2}\\
      \cdots\\
      \frac{\left[\sum_{i=1}^r (\lambda_{i+1}-\lambda_i)d_i\right]-\log(1-p)}{\lambda_{r+1}} & p\ge 1-e^{\left[\sum_{i=1}^r(\lambda_{i+1}-\lambda_{i})d_i\right]-\lambda_{r+1}d_r}.
  \end{cases}
\end{align*}
The \fct{qpwexp} function can calculate quantile values and \fct{rpwexp} function can generate PWE random variables from given parameters.

From a survival clinical trial, we observe $n$ pair of iid random variables $(T_1, \delta_1), (T_2, \delta_2), \cdots, (T_n, \delta_n)$, where $\delta_i$ indicates whether PWE distributed $X_i$ (i.e. an event of interest such as death) is observed ($\delta_i=1$) or not (non-informative right censored, $\delta_i=0$). Following \citet[Chpater 3.5]{klein2003survival}, the log-likelihood of the dataset can be constructed as:
\begin{align*}
    \log(L)=&\sum_{j\in D_1}\left[\log(\lambda_1)-\lambda_1 t_j\right]+\sum_{j\in C_1}\left[-\lambda_1 t_j\right]\\
    &\quad +\sum_{j\in D_2}\left[\log(\lambda_2)+(\lambda_2-\lambda_1)d_1-\lambda_2 t_j\right]+\sum_{j\in C_2}\left[(\lambda_2-\lambda_1)d_1-\lambda_2 t_j\right]\\
    &\quad +\cdots \\
    &\quad +\sum_{j\in D_{r+1}}\left\{\log(\lambda_{r+1}) +\sum_{i=1}^r \left[(\lambda_{i+1}-\lambda_i)d_i\right]-\lambda_{r+1} t_j\right\}+\sum_{j\in C_{r+1}}\left\{\sum_{i=1}^r \left[(\lambda_{i+1}-\lambda_i)d_i\right]-\lambda_{r+1} t_j\right\}
\end{align*}
where $D_k$ or $C_k$ represent the index set of event times or censoring times that fall into $k$th piece of the PWE distribution (i.e., $D_k=\{i\ |\ \delta_i=1, d_{k-1}\le T_i < d_k\}$, $C_k=\{i\ |\, \delta_i=0, d_{k-1}\le T_i < d_k\}$; here we let $d_0=0$, $d_{k+1}=+\infty$ for simpler notation). 
Obviously, the log-likelihood function $\log(L)$ is not smooth at $d_k=T_i$, resulting in maximum likelihood estimation infeasible using traditional optimization methods such as Newton-Raphson.

In the subsequent sections, we will first introduce the estimation of hazard rates when the positions of change-points are known (Section~\ref{sec:hazard}). Then we will move on to cases where we need to estimate both hazard rates and change-points without prior knowledge of positions of change-points (Section~\ref{sec:unknown}). Finally, we will cover the estimation of hazard rates and change-points when partial change-points are known (Section~\ref{sec:partial}).

\subsection{Parameter Estimation with Known Change-points}\label{sec:hazard}
In this section, we assume all change-points $d_1, d_2, \cdots d_r$ are already known. The log-likelihood function $\log(L)$ becomes smooth with respect to parameter parameter $\lambda$, allowing us to take a derivative of $\log(L)$ wrt $\lambda$:
\begin{align*}
    \frac{\partial \log(L)}{\partial \lambda_1}=&\frac{n_{D_1}}{\lambda_1}-\sum_{j\in C_1}t_j- n_{2^+}d_1\\
    \frac{\partial \log(L)}{\partial \lambda_2}=&\frac{n_{D_2}}{\lambda_2}-\sum_{j\in D_2, C_2}(t_j-d_1)- n_{3^+}(d_2-d_1)\\
    &\cdots\\
    \frac{\partial \log(L)}{\partial \lambda_r}=&\frac{n_{D_r}}{\lambda_r}-\sum_{j\in D_r, C_r}(t_j-d_{r-1})- n_{r+1^+}(d_r-d_{r-1})\\
    \frac{\partial \log(L)}{\partial \lambda_{r+1}}=&\frac{n_{D_{r+1}}}{\lambda_{r+1}}-\sum_{j\in D_{r+1}, C_{r+1}}(t_j-d_{r})
\end{align*}
where $n_{D_k}$ is the number of events that fall into the $k$th piece (i.e., $n_{D_k}=|D_k|$); $n_{k^+}$ is the number of events and censoring that fall into the $k$th to the last pieces (i.e., $n_{k^+}=\sum_{l=k}^{r+1}|D_l|+|C_l|$). 

Setting all derivatives equal to $0$, we obtain the maximum likelihood estimators of hazard rates:
\begin{align*}
    {\hat \lambda}_1=&\frac{n_{D_1}}{\sum_{j\in C_1}t_j+ n_{2^+}d_1}\\
    {\hat \lambda}_2=&\frac{n_{D_2}}{\sum_{j\in D_2, C_2}(t_j-d_1)+ n_{3^+}(d_2-d_1)}\\
    &\cdots \\
    {\hat \lambda}_r=&\frac{n_{D_r}}{\sum_{j\in D_r, C_r}(t_j-d_{r-1})+ n_{r+1^+}(d_r-d_{r-1})}\\
    {\hat \lambda}_{r+1}=&\frac{n_{D_{r+1}}}{\sum_{j\in D_{r+1}, C_{r+1}}(t_j-d_r)}.
\end{align*}

\subsection{Parameter Estimation with Unknown Change-points}\label{sec:unknown}
The log-likelihood function $\log(L)$ is differentiable with respect to $d_k$ as long as $d_k$ is not equal to any event time or censoring time ($d_k\ne T_i$). Under this assumption, we take the derivative of $\log(L)$ wrt $d_k$:
\begin{align*}
    \frac{\partial \log(L)}{\partial d_k}=&\sum_{j\in D_{k+1}}\left[\lambda_{k+1}-\lambda_k\right]+\sum_{j\in C_{k+1}}\left[\lambda_{k+1}-\lambda_k\right]\\
    =&n_{k+1^+}(\lambda_{k+1}-\lambda_k).
\end{align*}
Since $\lambda_{k+1}\ne \lambda_k$, then the expression above cannot be zero when $d_k$ falls between any two consecutive sample time points. This fact implies that only when $d_k$ is equal to any of the sample values $T_i$, the log-likelihood function achieves the maximum value \citep{loubert1986inference}. Based on these findings, we propose three estimation approaches in the following sections.

\begin{algorithm}[!t]
    \caption{Brute-force Search}\label{alg:change-point} 

    \begin{algorithmic}

        \Procedure{BFS}{times $T$, censoring status $C$, number of change-points $nbreak$, max number of combinations $max\_set$}
   \State $T\_sub \gets \textbf{sub\_sample}(T, nbreak, max\_set)$
   \State $candidate \gets combn(T\_sub, nbreak)$
   \State $candidate \gets candidate[, sample(ncol(candicate, max\_set))]$
   \For{$chg\_pt=$ each column in $candidate$}
   \State estimate hazard rates by the formula in Section~\ref{sec:hazard}
   \If{any hazard rate is $0$} 
    \State skip to next 
  \EndIf 
   \State calculate log-likelihood $L$ by the formula in Section~\ref{sec:loglike}
   \EndFor
   \State change-points $chg\_pt$ and hazard rates with largest $L$ is the MLE
   \EndProcedure\\

   \Function{sub\_sample}{$T$, $nbreak$, $max\_set$}
  \State \textbf{initialize} $nl \gets 1$
  \State \phantom{\textbf{initialize}} $nr \gets length(T)$
  \Repeat
  
  \State $NN \gets choose(floor((nl+nr)/2), nbreak)$

  \If{$NN > max\_set$} 
    \State $nr \gets floor((nl+nr)/2)$
  \Else
  \State $nl \gets floor((nl+nr)/2)$
  \EndIf 
  \Until{$nr - nl < 1.5$}

  \Return $T\_sub \gets sample(T, nr)$
  \EndFunction
    \end{algorithmic}
\end{algorithm}

\subsubsection{MLE by Brute-force Search} Inspired by \citet{qian2013multiple}'s least squared approach with grid search, we propose employing a brute-force search to calculate the log-likelihood values for all potential change-point combinations. For each change-point combination candidate, hazard rates can be estimated using the formula outlined in Section~\ref{sec:hazard} and thus the log-likelihood can be obtained. The change-points combination and hazard rates with the largest log-likelihood value are selected as the maximum likelihood estimator. The full brute-force search method guarantees to find the optimal solution, but it comes with a high computational burden.

Especially, when the number of samples or the number of change-points are relatively large, the combination of change-points becomes prohibitively large for a brute-force search to be feasible. In such instances, we draw random sub-samples first and then conduct an exhaustive search based on these sub-samples. The  \code{max\_set} argument in the \fct{pwexp.fit} function controls the maximum number of combinations to be explored. We summarize the entire procedure in pseudo-algorithm~\ref{alg:change-point}. This brute-force search with a sub-sampling method reduces computational intensity but introduces a risk of missing the optimal solution due to the sub-sampling process (i.e., the randomly selected sub-samples missed or do not cover the `optimal' change-points).

\subsubsection{OLS on Survival Function}
The second method for estimating change-points relies on the survival curve. 
By taking a logarithmic transformation of the PWE survival function, we obtain a piecewise linear function:
\begin{align*}
    \log(S(t))=&\begin{cases}
        {-\lambda_1 t}, & t<d_1\\
        {(\lambda_2-\lambda_1)d_1-\lambda_2t} , & d_1\le t<d_2\\
        \cdots\\
        {\left[\sum_{i=1}^r(\lambda_{i+1}-\lambda_{i})d_i\right]-\lambda_{r+1}t}, & t\ge d_r
    \end{cases}\\
    =&-\lambda_1 t +(\lambda_1-\lambda_2) (t-d_1)_+\cdots +(\lambda_r-\lambda_{r+1})(t-d_r)_+.
\end{align*}

Following \citet{kuchenhoff1996exact} and \citet{muggeo2003estimating}, we can fit a piecewise linear regression on log survival function $\log(S(t))$ with the \pkg{segmented} package and obtain the OLS estimation of change-points. Once the change-points are determined, hazard rates can be estimated using the formula described in Section~\ref{sec:hazard}. We summarize the entire procedure in pseudo-algorithm~\ref{alg:OLS}. Note that the OLS method does not obtain maximum likelihood estimators.

\begin{algorithm}[!h]
    \caption{OLS on Survival Function}\label{alg:OLS}
    \begin{algorithmic}
    \Procedure{OLS}{times $T$, censoring status $C$, number of change-points $nbreak$}
   \State $S \gets survfit(surv(T, C)\sim 1)$
   \State $chg\_pt \gets segmented(lm(log(S)\sim T))$
   \State estimate hazard rates by the formula in Section~\ref{sec:hazard}
   \State calculate log-likelihood $L$ by the formula in Section~\ref{sec:loglike}
   \EndProcedure
    \end{algorithmic}
\end{algorithm}

\subsubsection{Hybrid Method}
The change-points estimated in the OLS method are typically close to, but not exactly, the maximum likelihood estimators.   To obtain the true maximum likelihood estimator, we propose a hybrid method that combines the brute-force search with the OLS method.

\begin{algorithm}[!t]
  \caption{Hybrid Method}\label{alg:hybrid}
  \begin{algorithmic}
      \Procedure{HB}{times $T$, censoring status $C$, number of change-points $nbreak$, max number of combinations $max\_set$}
      \State $S \gets survfit(surv(T, C)\sim 1)$
      \State $chg$ with SE $\gets segmented(lm(log(S)\sim T))$
      \State $chg\_candidate_1 \gets T$ in $95\%$ CI of $chg_1$
      \State $\vdots$
      \State $chg\_candidate_{nbreak} \gets T$ in $95\%$ CI of $chg_{nbreak}$
      \State $candidate \gets expand.grid(chg\_candidate_1, \cdots, chg\_candidate_{nbreak})$
      \State $candidate \gets candidate[sample(nrow(candicate, max\_set)),]$
 \For{$chg\_pt=$ each row in $candidate$}
 \State estimate hazard rates by the formula in Section~\ref{sec:hazard}
 \If{any hazard rate is $0$} 
  \State skip to next 
\EndIf 
 \State calculate log-likelihood $L$ by the formula in Section~\ref{sec:loglike}
 \EndFor
 \State change-points $chg\_pt$ and hazard rates with largest $L$ are returned
 \EndProcedure
  \end{algorithmic}
\end{algorithm}

Specifically, instead of randomly drawing sub-samples in brute-force search, we draw sub-samples from the values near the estimated change-points from the OLS method.
For example, we may consider samples within some percentage of the confidence intervals around the change-points.  Subsequently, we conduct an exhaustive search based on these sub-samples. 
The  \code{max\_set} argument in the \fct{pwexp.fit} function controls the maximum number of combinations to be explored. This hybrid method is significantly more efficient and has a high probability of finding the optimal solution in practice. Based on our experience, this approach fits most datasets well and overcomes previous limitations of both Brute-force Search and OLS methods. The entire procedure of hybrid method is summarized in pseudo-algorithm~\ref{alg:hybrid}.

\subsection{Parameter Estimation with Partial Known Change-points}\label{sec:partial}
The proposed \pkg{PWEXP} package also accepts partial known change-points when some of the change-points are specified by the \code{breakpoint} argument in the \fct{pwexp.fit} function and the total number of change-points (\code{nbreak}) is larger than the length of \code{breakpoint}. This feature is particularly useful when the user wants to manually include certain change-points in the fitted model while retaining the flexibility of the PWE model. For example, if the survival curve is expected to change at the start of a  treatment switch, we can manually specify a change-point at this time point. To achieve this: 

\begin{itemize}
    \item For the brute-force search method, the user will conduct an exhaustive search on the combinations of \code{length(breakpoint)} pre-specified change-points along with (\code{nbreak - length(breakpoint)}) candidate change-points from data to obtain the maximum likelihood estimator.  
    \item For the OLS on survival function method, when fitting a piecewise linear regression on log survival function $\log(S(t))$ with the \fct{segmented} function from the \pkg{segmented} package, the user can set the \code{npsi} argument to the number of unknown change-points (\code{npsi = nbreak - length(breakpoint)}) and the \code{fixed.psi} argument to the pre-specified change-points (\code{fixed.psi = breakpoint}). Then the fitted result will have a total of \code{nbreak} change-points including all pre-specified change-points. 
    \item For the hybrid method, the user first follows the approach used in the OLS on survival function method to identify  the remaining (\code{nbreak - length(breakpoint)}) change-points. Subsequently, the user can fix the known change-points (specified  in \code{breakpoint}) and draw sub-samples from the values near the estimated change-points from the OLS method. Finally, the user can conduct an exhaustive search based on these sub-samples to obtain the maximum likelihood estimator. 
\end{itemize}

\subsection{Parameter Estimation Summary}
\begin{figure}[!h]
\centering
\includegraphics[width=.9\textwidth]{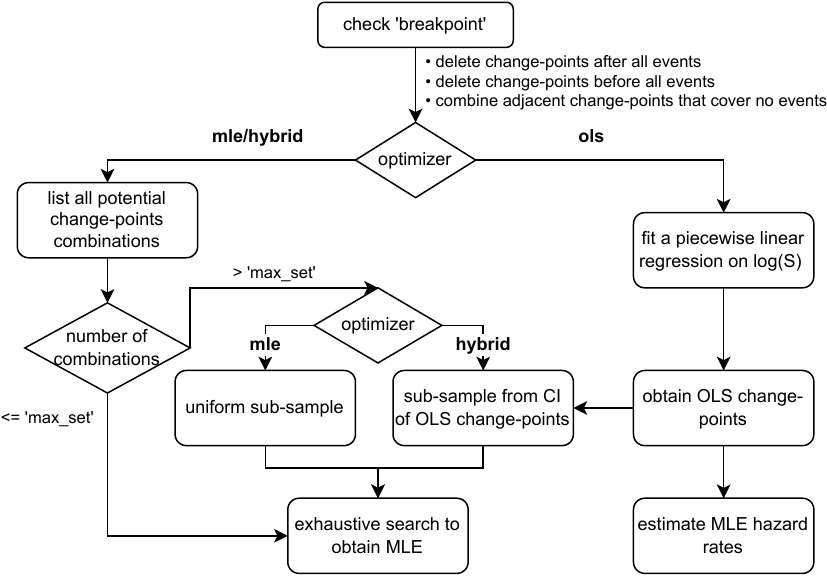}
\caption{Diagram of the process for estimating a PWE model using the \fct{pwexp.fit} function}\label{fig:estimation}
\end{figure}

The \fct{pwexp.fit} function is designed for PWE model estimation. 
Figure~\ref{fig:estimation} illustrates the process for estimating a PWE model: The \fct{pwexp.fit} function first validates the \code{breakpoint} argument (if not \code{NULL}). Any change-points that are too early or too late such that there are no events before or after them are deleted with a warning. If two adjacent change-points are too close to each other such that there are no events between them, they are combined by taking their average. Subsequently, based on the \code{optimizer} argument, the function performs parameter estimation as described in the preceding sections.

To quantify the uncertainty of estimated model parameters, we can employ the \fct{boot.pwexp.fit} function to conduct bootstrapping. This technique involves resampling from the original dataset with replacement to generate multiple bootstrap samples. For each resampled dataset, we fit the PWE model and record the parameter estimates. By analyzing the distribution of parameter estimates obtained from the bootstrap samples, we can construct confidence intervals to capture the uncertainty associated with the model parameters. 

\subsection{Tail Robustness}\label{sec:tail}
\begin{figure}[!h]
\centering
\includegraphics[clip, trim=0 15 0 37pt, width=.7\textwidth]{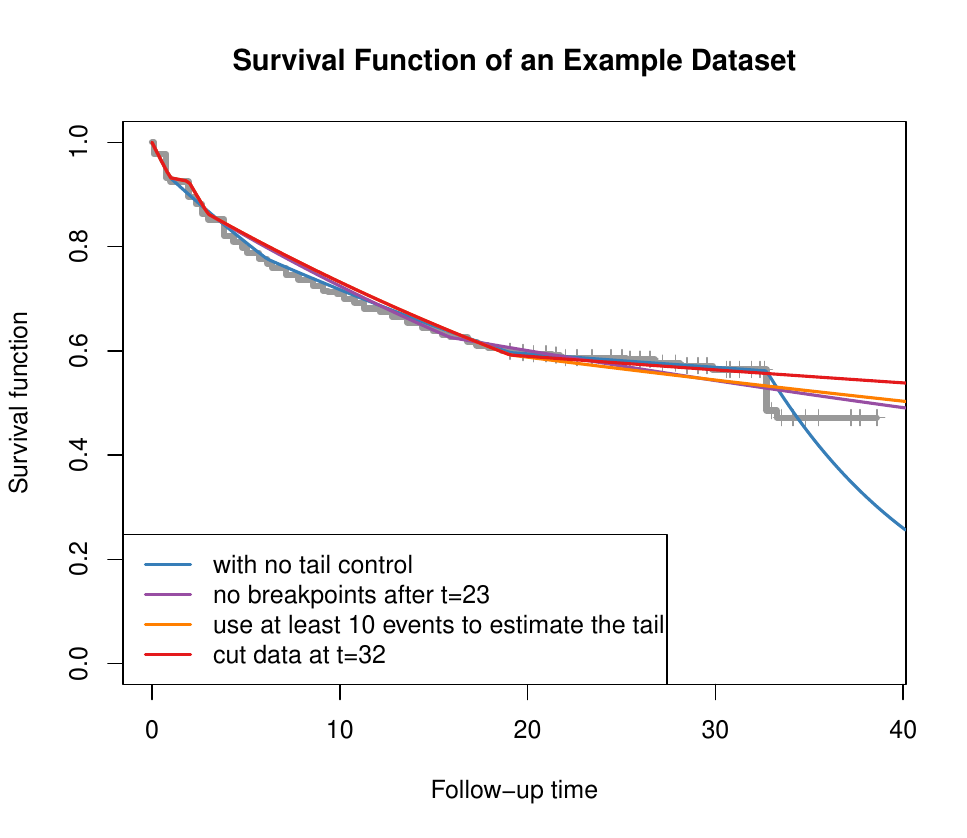}
\caption{Survival function of an example dataset. This is an illustration of different approaches to adjust tail.}\label{fig:tail}
\end{figure}

In a real-world dataset, some `abnormal' patterns such as sudden drops  may occur in the tail of a survival curve.  For example, in Figure~\ref{fig:tail}, the gray Kaplan-Meier curve exhibits a sudden drop at $t=32$. If this abrupt change is not properly addressed and a piecewise exponential model is fitted directly to the data, it will result in the blue curve, which over-estimates the hazard rate in the tail.

To mitigate this issue, several strategies can be employed:

\begin{enumerate}
  \item One strategy is to ensure that a sufficient number of events are used in estimating the hazard rate for the tail of the survival curve.  This can be achieved by setting a minimum threshold for the number of events required in the estimation of the last exponential piece. The \code{min\_pt\_tail} argument (default: \code{min\_pt\_tail = 5}) in the \fct{pwexp.fit} function is used to specify this minimum number.  
  For example, setting \code{min\_pt\_tail = 10}, we will have at least $10$ events to be used to estimate the tail, resulting in a more robust estimation demonstrated by the orange curve in Figure~\ref{fig:tail}.
  \item Another strategy is to constrain the position of the last change-point to prevent it from occurring too late. This actually increases the number of events used in estimating the last exponential piece as well.
  The \code{exclude\_int} argument in the \fct{pwexp.fit} function can be used to exclude an interval  from having any change-points. For example, by setting \code{exclude\_int = c(23, Inf)}, we exclude any change-points after time $t=23$, leading to the purple curve in Figure~\ref{fig:tail}.
  \item The third strategy involves cutting (re-censoring) the data prior to the occurrence of the sudden drop by using the \fct{cut\_dat} function. Subsequently, a PWE model is fitted to the truncated data. For example, here we re-censor the data at $t=32$ and red curve is the fitted model. 
\end{enumerate}

These strategies offer practical solutions to address `abnormal' tails in survival data and enhance the reliability of hazard rate estimations.

\subsection{Determination of Change-points Numbers}\label{sec:number}
In the preceding sections, the number of change-points is assumed to be predetermined. However, in practice, the optimal number of change-points is usually unknown. 
One common approach is to plot the KM curve and fit a series of PWE models, starting from $0$ change-points and gradually increasing the number of change-points. 
By performing a virtual comparison between the KM curve and the fitted PWE models, to ascertain the adequacy of the number of change-points in the PWE model for accurately fitting the KM curve.

\sloppy Furthermore, statistical criteria such as Akaike Information Criterion (AIC) (from the \fct{pwexp.fit} function), Bayesian Information Criterion (BIC) (from the \fct{pwexp.fit} function), and cross-validation log-likelihood (from the \fct{cv.pwexp.fit} function) can assist in determining the optimal number of change-points. These statistical metrics provide a more objective basis for assessing `goodness-of-fit' of the model, and help to select the most appropriate model configuration.

\section{Event Prediction with PWE}\label{sec:prediction}
In this section, we will discuss the process of conducting event prediction using the PWE model. Prior to delving into the details, we need some preliminary knowledge outlined in Section~\ref{sec:con}. Following this, we will elaborate on event prediction methodologies.

\subsection{Conditional PWE Distribution}\label{sec:con}
Suppose random variable $T$ follows a PWE distribution as described in Section~\ref{sec:loglike}, then the conditional survival function ($1-$cumulative density function) and cumulative distribution function of $T$ given $T>R$ is given by 
\begin{align*}
    S(t|t>R)&=\frac{S(t)}{S(R)}, \quad \text{where } S(t), S(R) \text{ as defined in Section~\ref{sec:loglike}}\\
    F(t|t>R)&=1-\frac{S(t)}{S(R)}, \quad \text{where } S(t), S(R) \text{ as defined in Section~\ref{sec:loglike}}.
\end{align*}
The \fct{conditional\_ppwexp} function in the \pkg{PWEXP} package can be used to calculate these values.

The corresponding conditional PWE random variable can be generated with inverse transform method, which relies on the conditional quantile function $Q(p|t>R)$:
\begin{align*}
    Q(p|t>R)=\begin{cases}
        \frac{\lambda_1 R-\log(1-p)}{\lambda_{1}}, & p <F(d_{1}|t>R), R <d_{1}\\
        \frac{\left[\sum_{i=1}^{k-1} (\lambda_{i+1}-\lambda_i)d_i\right]+\lambda_1 R-\log(1-p)}{\lambda_{k}}, & F(d_{k-1}|t>R)\le p <F(d_{k}|t>R), R <d_{1}\\
        \cdots \\
        \frac{\left[\sum_{i=m}^{k-1} (\lambda_{i+1}-\lambda_i)d_i\right]+\lambda_m R-\log(1-p)}{\lambda_{k}}, & F(d_{k-1}|t>R)\le p <F(d_{k}|t>R), d_{m-1}\le R <d_{m}
    \end{cases}
\end{align*}

\sloppy The \fct{conditional\_qpwexp} function can calculate conditional quantile values and the \fct{conditional\_rpwexp} function can generate the conditional PWE random variables from given parameters.

\subsection{Event Prediction}\label{sec:pred}
As discussed in Section~\ref{sec:intro}, following the classic event prediction framework \citep{bagiella2001predicting, ying2008weibull}, at the current calendar time, $t_0$, the conditional expectation of the number of events $ED(t_0,t')$ by time $t'$ ($t'>t_0$) is:
\begin{align*}
  ED(t_0,t')=D(t_0)+Q(t_0,t')+R(t_0,t'),
\end{align*} 
where $D(t_0)$ is the observed events by $t_0$; $Q(t_0, t')$ is the expected number of subjects who are in the study with no events by time $t_0$ and will have events by $t'$; and $R(t_0,t')$ is the expected number of subjects who will enroll and have events between $t_0$ and $t'$. 

Assume that the PWE distributed event times $T_i$ are independent, and that the PWE distributed censoring times $C_i$ are also independent of each other and of the event times.

Suppose $R$ is the index set of the subjects who will enroll at $U_i > t_0$. For an individual subject, the probability of observing an event before $t'$ is:
\begin{align*}
  P^R_i=\Pr(Y_i + U_i \le t', T_i < C_i)=\Pr(\min(T_i, C_i) + U_i \le t', T_i < C_i).
\end{align*}
Define random variable $V_l=I(\min(T_l, C_l) + U_i \le t, T_l < C_l)$, where $T_l$ and $C_l$ can be generated as described in Section~\ref{sec:loglike} using the \fct{rpwexp} function. Then $R(t_0,t')$ can be approximately simulated as
\begin{align*}
  R(t_0,t')=\sum_{i\in R}P^R_i\approx \sum_{i\in R} \left\{\frac1N\sum_{l=1}^N V_l \right\},
\end{align*} 
where $N$ is the number of simulation iterations controlled by the \code{n\_each} argument in the \fct{predict} function of the \pkg{PWEXP} package. 

Suppose $Q$ is the index set of the subjects who  are in the study with no events at time $t_0$. For an individual subject, the probability of observing an event before $t'$ is:
\begin{align*}
  P^Q_i=&\Pr(Y_i + U_i \le t', T_i < C_i | Y_i + U_i > t_0)\\
  =&\Pr(\min(T_i, C_i) + U_i \le t', T_i < C_i | \min(T_i, C_i) + U_i > t_0)\\
  =&\Pr(\min(T_i, C_i) + U_i \le t', T_i < C_i | T_i + U_i > t_0, C_i + U_i > t_0).
\end{align*}
Define $W_l=I(\min(T_l, C_l) + U_i \le t', T_l < C_l) | T_l > t_0- U_i, C_l  > t_0- U_i$, where the conditional $T_l$ and $C_l$ can be generated as described in Section~\ref{sec:con} using the \fct{conditional\_rpwexp} function. Then $Q(t_0,t')$ can be approximately simulated as 
\begin{align*}
  Q(t_0,t')=\sum_{i\in Q}P^Q_i\approx\sum_{i\in Q}\left\{\frac1N\sum_{l=1}^N W_l \right\},
\end{align*} 
where $N$ is the number of simulation iterations controlled by the \code{n\_each} argument in the \fct{predict} function. 

Bringing all these elements together, the expectation of the number of events $ED(t_0,t')$ can be simulated as 
\begin{align*}
  ED(t_0,t')\approx D(t_0)+\sum_{i\in Q}\left\{\frac1N\sum_{l=1}^N W_l \right\}+\sum_{i\in R} \left\{\frac1N\sum_{l=1}^N V_l \right\}.
\end{align*} 

Similarly, we can define the predictive number of events $PED(t_0,t')$ as the result of a single generation:
\begin{align*}
  PED(t_0, t')=D(t_0)+\sum_{i\in R}  V_l +\sum_{i\in Q} W_l.
\end{align*}
Both $ED(t_0,t')$ and $PED(t_0,t')$ can be estimated using the \fct{plot\_event} function. 

\subsection{Confidence Interval}
We will adopt percentile bootstrap to compute  the confidence interval of the expected number of events $ED(t_0, t')$.  The procedure comprises several steps. First we perform case resampling (with replacement) to obtain \code{nsim} samples. Then we conduct \code{nsim} PWE model estimations to get \code{nsim} sets of parameters. For each set of parameters, we estimate the bootstrapped $ED(t_0, t')^*$ following the method outlined in the preceding section. Finally, 
\begin{align*}
  \left( ED(t_0, t')^*_{(\alpha/2)}, ED(t_0, t')^*_{(1-\alpha/2)} \right)
\end{align*}
represents the $1-\alpha/2$ percentile of the bootstrapped $ED(t_0, t')^*$.

For a predictive interval of the number of events $PED(t_0, t')$, we will follow a similar procedure.  We conduct case resampling (with replacement) to obtain \code{nsim} samples and conduct PWE model estimations to obtain  \code{nsim} sets of parameters. For each set of parameters, we repeatedly generate  $N$ (the \code{n\_each} argument in the \fct{predict} function) predictive number of event $PED(t_0, t')^*$ as described in the last section. Finally, 
\begin{align*}
  \left( PED(t_0, t')^*_{(\alpha/2)}, PED(t_0, t')^*_{(1-\alpha/2)} \right)
\end{align*}
denotes the $1-\alpha/2$ percentile of the bootstrapped $PED(t_0, t')^*$. The predictive interval accounts for  both the uncertainty stemming from model estimation and the variability of event occurrence. As a result, it is always wider than the confidence interval of the expected number of events $ED(t_0, t')$. The \fct{plot\_event} function can provide both types of confidence intervals. 


\section[Description of the PWEXP Package]{Description of the \pkg{PWEXP} Package} \label{sec:package}
\subsection{Installation}
The \pkg{PWEXP} package is developed in \proglang{R}
and available from CRAN at \url{https://CRAN.R-project.org/package=PWEXP}. Hence,
it can be installed in any computer with \proglang{R} with the command:
\begin{verbatim}
R> install.packages("PWEXP")
\end{verbatim}

The development version of the package can be installed from GitHub using \pkg{devtools} with the following command:
\begin{verbatim}
R> devtools::install_github("zjph602xtc/PWEXP")
\end{verbatim}
These commands also import \pkg{fastmatch} package and \pkg{segmented} package. Packages \pkg{parallel}, \pkg{doSNOW} and \pkg{foreach} are needed if multi-core calculation is enabled for bootstrapping and cross-validation. Several vignettes have been prepared using \pkg{knitr}. An online package tutorial is available at \url{https://zjph602xtc.github.io/PWEXP/}.

\subsection{Software Architecture and Main Functions}

\begin{figure}[!h]
  \includegraphics[width=\textwidth]{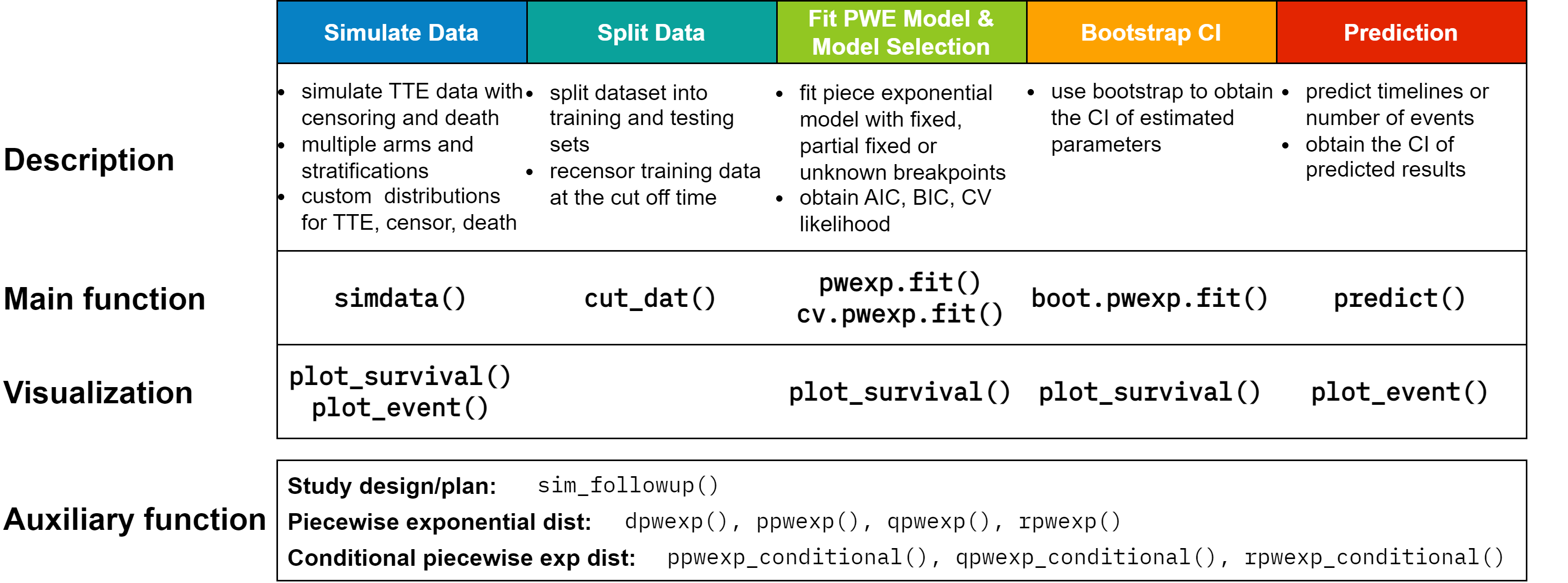}
  \caption{Main functions and structure of the \pkg{PWEXP} package}\label{fig:structure}
\end{figure}

The \pkg{PWEXP} package follows a software architecture based on functional programming.
Each main function returns an instance of an \code{S3} class, which implements relevant \code{S3} methods to facilitate downstream analyses. The \pkg{PWEXP} package offers a comprehensive suite of tools tailored for analyzing survival data with a piecewise exponential distribution. It facilitates various tasks related to survival analysis and event prediction based on the fitted model. Specifically, the package comprises several components designed for different tasks as shown in Figure~\ref{fig:structure}:

\begin{itemize}
  \item \textbf{Survival data simulation}
    \begin{itemize}
      \item the \fct{simdata} function generates simulated survival dataset with:
        \begin{itemize}
          \item randomization/enrollment time defined by randomization/enrollment rate (number of subjects per month) or randomization/enrollment curve (number of subjects in each month);
          \item multiple treatment groups with stratification by user-defined allocation ratio;
          \item primary endpoint (event), drop-out and death with exponential distribution or user-defined distributions (e.g., piecewise exponential, mixture distribution, etc.) for each stratification in each group.
        \end{itemize}
      \item the \fct{plot\_survival} function visualizes KM estimator for right censored data;
      \item the \fct{plot\_event} function plots the number of cumulative events.
    \end{itemize}
  \item \textbf{Data cut-off}
    \begin{itemize}
      \item the \fct{cut\_dat} function truncates the data by a clinical cut-off date (DCOD), only retaining subjects randomized before this time and re-censoring the data. It is useful to create a training data set.
    \end{itemize}
  \item \textbf{PWE model estimation}
    \begin{itemize}
      \item  the \fct{pwexp.fit} function fits the piecewise exponential model to right censored data with:
        \begin{itemize}
          \item pre-specified change-points;
          \item partially pre-specified change-points;
          \item unknown change-points (to be fitted from the data).
        \end{itemize}
      \item the \fct{pwexp.fit} function also calculates model AIC, BIC;
      \item the \fct{cv.pwexp.fit} conducts model cross-validation and obtains the CV log-likelihood;
      \item the \fct{plot\_survival} plots the fitted survival curve for \code{pwexp.fit} object.
    \end{itemize}
  \item \textbf{Model bootstrapping}
    \begin{itemize}
      \item the \fct{boot.pwexp.fit} conducts bootstrapping for an existing model;
      \item the \fct{plot\_survival} plots fitted survival curve and its  confidence interval  for \code{boot.pwexp.fit} object.
    \end{itemize}
  \item \textbf{Events/timeline prediction}
    \begin{itemize}
      \item  the \fct{predict} function  predicts the number of events and timeline based on the event and drop-out model (optional);
      \item the \fct{predict} function provides the CI of the predicted number of events or timeline for a bootstrapping model;
      \item  the \fct{plot\_event} plots predicted number of events (for \code{predict.pwexp.fit} object) and its CI (if applicable, for \code{predict.boot.pwexp.fit} object).
    \end{itemize}
  \item \textbf{Auxiliary functions}
    \begin{itemize}
      \item the auxiliary function \fct{sim\_followup} estimates follow-up time and number of events at design stage;
      \item  the auxiliary functions \fct{dpwexp}, \fct{ppwexp}, \fct{qpwexp}, \fct{rpwexp} are the PDF, CDF, quantile function, generator function for PWE distribution;
      \item  the auxiliary functions \fct{ppwexp\_conditional}, \fct{qpwexp\_conditional}, \fct{rpwexp\_conditional} are the CDF, quantile function, generator function for conditional PWE distribution.
    \end{itemize}
\end{itemize}

In the following sections, we will present two examples to illustrate the usage of the \pkg{PWEXP} package. The first example focuses on event prediction using simulated data, while the second example demonstrates a study design based on a real case study.

\section{Event/timeline Prediction Example}\label{sec:simulation}

In this section, we will present a comprehensive workflow for event and timeline prediction. The workflow comprises three steps. First, we simulate survival data and truncate the dataset at the time $t_0$ when $80\%$ of the subjects are randomized. Subsequently, we fit a PWE model to the truncated dataset. Finally, we perform event and timeline prediction for time $t'$ with $t'>t_0$ by using the estimators from the truncated dataset and compare the predicted events/timelines with the `real' data from simulation.

\subsection{Simulate PWE Data}
Here we utilize the \fct{simdata} function to create a simple example dataset with the following characteristics:
\begin{itemize}
  \item randomization rate is defined as $20$ subjects per month (\code{rand\_rate = 20}), and total sample size is $1000$ (\code{total\_sample = 1000});
  \item the primary endpoint (event)  follows a PWE distribution with monthly hazard rates of $0.1$ for $t < 5$ months, $0.01$ for $5 \le t < 14$ months, $0.2$ for $t \ge 14$ months (defined by \code{myevent\_dist} function);
  \item the drop-out follows an exponential distribution with a drop-out probability of $3\%$/month (\code{drop\_rate = 0.03})(equivalently, a monthly drop-out hazard rate is $-\log(1-0.03)=0.0304$);
  \item the \code{add\_column} argument requests additional variables to be included in the dataset.
\end{itemize}

\begin{verbatim}
R> myevent_dist <- function(n)rpwexp(n, rate = c(0.1, 0.01, 0.2), 
+                                    breakpoint = c(5,14))
R> dat <- simdata(rand_rate = 20, total_sample = 1000, drop_rate = 0.03,  
+                advanced_dist = list(event_dist = myevent_dist),
+                add_column = c('censor_reason', 'event', 'followT', 'followT_abs'))
R> head(dat)

  ID    randT     eventT     dropT censor_reason event    followT followT_abs
1  1 18.95190  0.9741066 47.025224          <NA>     1  0.9741066    19.92601
2  2 18.38245  3.1487676  1.336703      drop_out     0  1.3367033    19.71915
3  3 38.76302 14.3753607 35.737477          <NA>     1 14.3753607    53.13838
4  4 27.59118  3.2444541 12.549571          <NA>     1  3.2444541    30.83563
5  5 48.42573 16.1987174 24.916275          <NA>     1 16.1987174    64.62445
6  6 35.62735 14.1660874 62.585078          <NA>     1 14.1660874    49.79344
\end{verbatim}

In the generated dataset \code{dat}, all subjects are followed for infinite time and we will truncate it by a clinical cut-off date (DCOD) with the \fct{cut\_dat} function in the next step. These variables are included in dataset \code{dat} (See Appendix~\ref{app:simudata} for more details):
\begin{itemize}
  \item \code{randT} is the randomization time for each subject;
  \item \code{eventT} and \code{dropT} are the event time and drop-out time, respectively;
  \item \code{followT} is the follow-up time, which is the minimum value of \code{eventT}, \code{dropT}. In real-world datasets, this is the observation time;  
  \item \code{followT\_abs} is the sum of  \code{randT} and \code{followT};
  \item \code{event} indicates whether the primary event occurred at the end of follow-up with $0$ for censoring and $1$ for occurrence of event. If a subject is censored, \code{censor\_reason} shows the type of censoring (i.e., \code{drop\_out}, \code{death} or \code{never\_event} (when \code{followT == Inf})).
\end{itemize}

\begin{figure}[!h]
\parbox{0.5\textwidth}{
\textbf{(a)}\\
\includegraphics[width=0.5\textwidth]{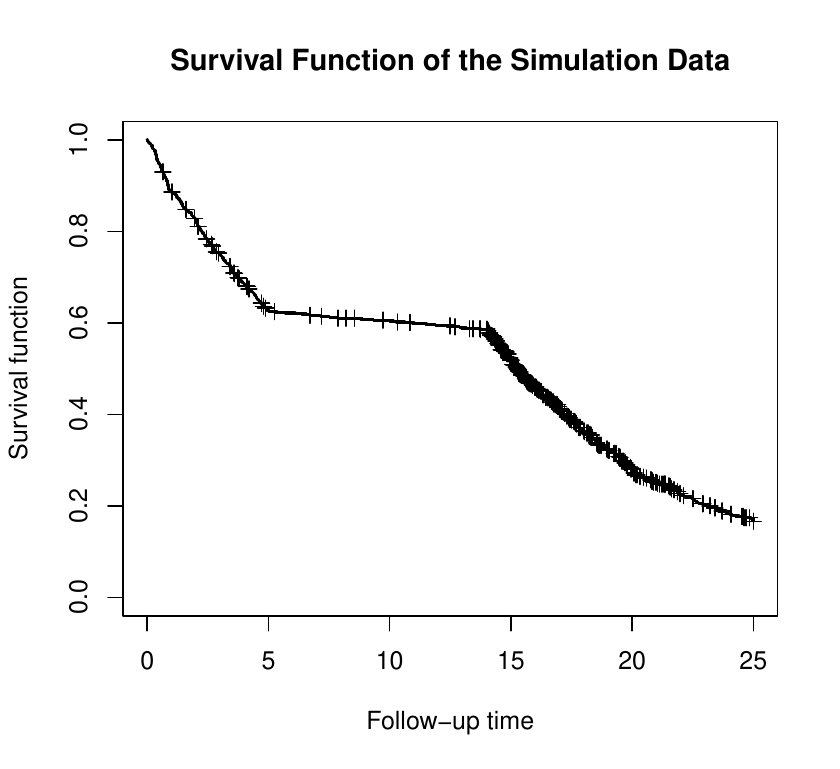}
}
\parbox{0.5\textwidth}{
\textbf{(b)}\\
\includegraphics[width=0.5\textwidth]{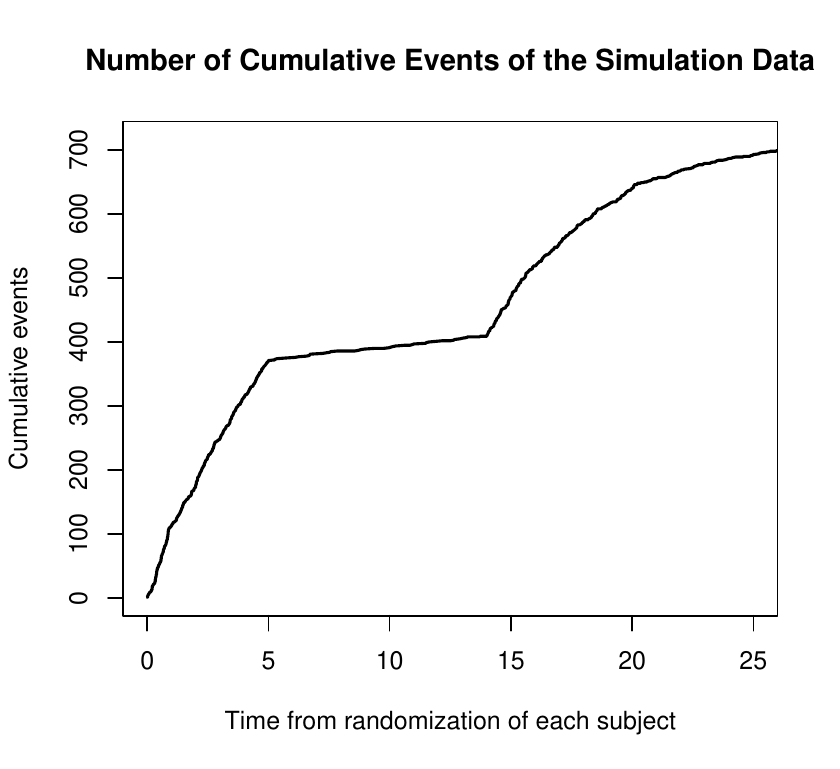}
}
\caption{(a) KM curve for the simulated dataset; (b) Event curve for the simulated dataset.}\label{fig:demo1}
\end{figure}

We can visualize the survival curve and cumulative number of events using the \fct{plot\_survival}, \fct{plot\_event} functions respectively. Figure~\ref{fig:demo1}(a) displays the KM curve of the simulated events. 
Notably, the curve drops faster before $t=5$ and after $t=14$ with hazard rates being $0.1$ and $0.2$, respectively, which are larger than the hazard rate of $0.01$ in the middle time interval.  Figure~\ref{fig:demo1}(b) is the cumulative number of events over time. 
It is important to note that the x-axis represents the time from randomization of each subject, allowing us to observe changes in the event rate due to the piecewise exponential distribution (by setting argument \code{abs\_time = F} in the the \fct{plot\_event} function). If we were to plot the event curve from the beginning of the trial, the figure would be Figure~\ref{fig:pred1}.

\begin{verbatim}
R> plot_survival(dat$eventT, dat$event, conf.int = F, mark.time = T, 
+               xlim = c(0, 25), main = 'Survival Function of the Simulation Data')
R> plot_event(dat$eventT, abs_time = F, dat$event, xlim = c(0, 25), 
+            main = 'Number of Cumulative Events of the Simulation Data')
\end{verbatim}

To compare the prediction results with the ground truth, we need to cut the complete dataset \code{dat} at the interim analysis time. For instance, we can cut the dataset using the time of the clinical cut-off date (DCOD) when $80\%$ of subjects are randomized. This operation retains only data available by DCOD in the returned data \code{train}. Additionally, the returned data will be re-censored at the DCOD. 

\begin{verbatim}
R> cut <- quantile(dat$randT, probs = 0.8)
R> cut

   80% 
39.99107 

R> train <- cut_dat(cut = cut, data = dat, var_randT = 'randT', 
+                  var_followT = 'followT', var_followT_abs = 'followT_abs',
+                  var_event = 'event', var_censor_reason = 'censor_reason')
R> head(train)

  ID    randT     eventT     dropT censor_reason event   followT followT_abs
1  1 18.95190  0.9741066 47.025224          <NA>     1 0.9741066    19.92601
2  2 18.38245  3.1487676  1.336703      drop_out     0 1.3367033    19.71915
3  3 38.76302 14.3753607 35.737477           cut     0 1.2280545    39.99107
4  4 27.59118  3.2444541 12.549571          <NA>     1 3.2444541    30.83563
6  6 35.62735 14.1660874 62.585078           cut     0 4.3637258    39.99107
7  7 13.92114  0.9511499 85.247486          <NA>     1 0.9511499    14.87229
\end{verbatim}

\subsection{PWE Model Estimation}
In this section, we aim to determine the optimal piecewise exponential model for the training dataset \code{train}. Given that the position and number of change-points are unknown in practice, we will fit models \code{fit\_b0} to \code{fit\_b4} with $0$ to $4$ unknown change-points. The \code{nbreak} argument is the total number of change-points. It is worth noting that the \code{fit\_b0} model is actually an exponential model since there are no change-points. 

\begin{verbatim}
R> fit_b0 <- pwexp.fit(train$followT, train$event, nbreak = 0)
R> fit_b1 <- pwexp.fit(train$followT, train$event, nbreak = 1)
R> fit_b2 <- pwexp.fit(train$followT, train$event, nbreak = 2)
R> fit_b3 <- pwexp.fit(train$followT, train$event, nbreak = 3)
R> fit_b4 <- pwexp.fit(train$followT, train$event, nbreak = 4)
\end{verbatim}
As indicated in Section~\ref{sec:number}, the initial step is to visualize the fitted curves alongside  the real KM curve. Now let's proceed by plotting the fitted curves:

\begin{verbatim}
R> plot_survival(train$followT, train$event, xlim = c(0, 40), 
+                main = 'Fitted Models with Different Number of Breakpoints')
R> plot_survival(fit_b0, col = 'green', lwd = 2)
R> plot_survival(fit_b1, col = 'blue', lwd = 3, show_breakpoint = F)
R> plot_survival(fit_b2, col = 'red', lwd = 3, 
+                breakpoint_par = list(col = 'grey50', lty = 2))
R> plot_survival(fit_b3, col = 'orange', lwd = 3, show_breakpoint = F)
R> legend('topright', c('training data', '0 breakpoints', '1 breakpoint',
+                       '2 breakpoints', '3 breakpoints'), lwd = 3, 
+         col = c('black', 'green', 'blue', 'red', 'orange'))
\end{verbatim}

\begin{figure}[!h]
\centering
\includegraphics[clip, trim=0 16 0 37pt, width=.7\textwidth]{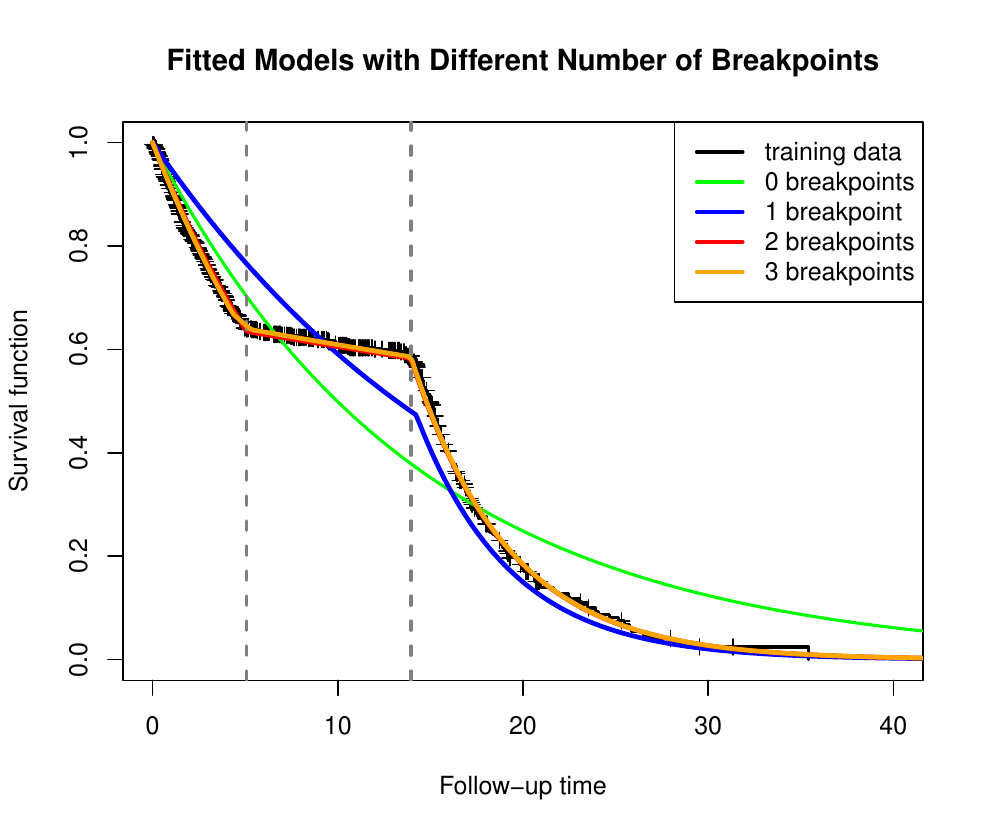}
\caption{Visualization of PWE models with different number of change-points}\label{fig:curve}
\end{figure}

Figure~\ref{fig:curve} shows the four fitted models. It is clear that the exponential model does not adequately capture the observed survival patterns. However, the PWE model with one change-point, located around $t=14$, demonstrated a much closer alignment with the actual KM curve.  Moreover, the PWE model with two change-points exhibits a perfect fit. Increasing the number of change-points beyond two does not yield significant improvement in the model fit. 

Visualization method can be combined with statistical criteria such as AIC, BIC, and CV Log likelihood to aid in the process of model selection. For illustrative purposes, we fit two additional models:
\begin{enumerate}
  \item \code{fit\_a0} with two pre-specified change-points at $t=5$ and $t=14$;
  \item \code{fit\_a1} with one pre-specified change-point  at $t=14$, while the total number of change-points is set to $2$.  The position of the other change-point will be estimated from the training data.
\end{enumerate}
\begin{verbatim}
R> fit_a0 <- pwexp.fit(train$followT, train$event, breakpoint = c(5,14))
R> fit_a1 <- pwexp.fit(train$followT, train$event, nbreak = 2, breakpoint = c(14))
\end{verbatim}
Note that models \code{fit\_a0} and \code{fit\_a1} are solely for demonstration purposes and will not be considered for subsequent analysis, as the actual positions of change-points will not be known in practice.

\begin{figure}[!h]
\centering
\includegraphics[clip, trim=0 17 0 37pt, width=.7\textwidth]{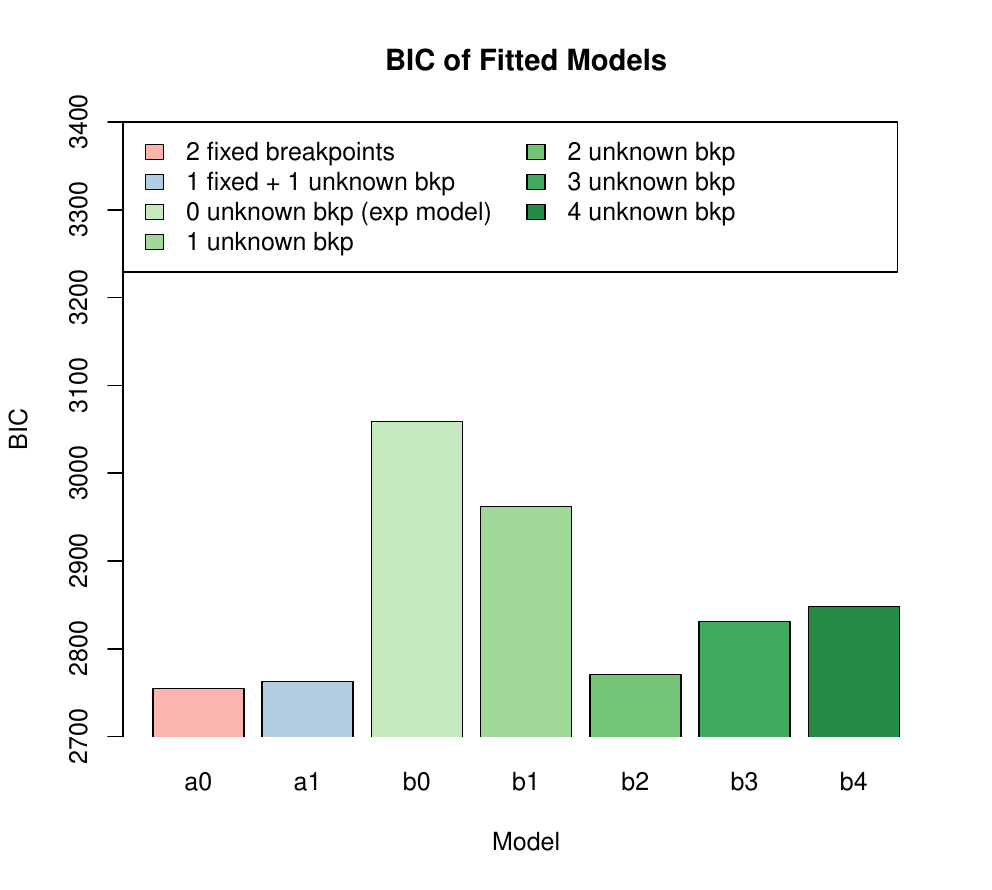}
\caption{BIC of fitted PWE models}\label{fig:bic}
\end{figure}

The BIC values for the six models are plotted in Figure~\ref{fig:bic}.  The model \code{fit\_a0} with two pre-specified true known change-points has the smallest BIC and serves as a baseline. 
Among the models without pre-specified change-points, we observe that \code{fit\_b2}, which has the correct number of change-points, exhibits the lowest BIC value.

\begin{verbatim}
R> model_name <- c('2 fixed breakpoints', '1 fixed + 1 unknown bkp', 
+                  '0 unknown bkp (exp model)', '1 unknown bkp', '2 unknown bkp', 
+                  '3 unknown bkp', '4 unknown bkp')
R> model_ind <- c("a0", "a1", "b0", "b1", "b2", "b3", "b4")
R> col <- c("#FBB4AE", "#B3CDE3", "#C7E9C0", "#A1D99B", "#74C476", "#41AB5D", "#238B45")

R> barplot(c(fit_a0$BIC, fit_a1$BIC, fit_b0$BIC, fit_b1$BIC, fit_b2$BIC, 
+            fit_b3$BIC, fit_b4$BIC), xlab = 'Model', 
+         ylab = 'BIC', ylim = c(2700, 3400), xpd = F, col = col, 
+         names.arg = model_ind, main = 'BIC of Fitted Models')
R> legend('topleft', model_name, fill = col, ncol = 2, text.width = 3.5)
\end{verbatim}

Cross-validation (CV) log-likelihood can be used to further assist in model selection. The \fct{cv.pwexp.fit} function conducts cross-validation either from an estimated PWE model or directly from the data. Additionally, the \code{parallel} argument enables parallel computing to reduce the computation time.
\begin{verbatim}
R> fit_a0_cv <- cv.pwexp.fit(fit_a0, nsim = 100)
R> fit_a1_cv <- cv.pwexp.fit(fit_a1, nsim = 100)
R> fit_b0_cv <- cv.pwexp.fit(fit_b0, nsim = 100)
R> fit_b1_cv <- cv.pwexp.fit(fit_b1, nsim = 100)
R> fit_b2_cv <- cv.pwexp.fit(fit_b2, nsim = 100, parallel = TRUE, mc.core = 10)
R> fit_b3_cv <- cv.pwexp.fit(fit_b3, nsim = 100, parallel = TRUE, mc.core = 10)
R> fit_b4_cv <- cv.pwexp.fit(train$followT, train$event, nbreak = 4, nsim = 100)
\end{verbatim}

\begin{figure}[!h]
\centering
  \includegraphics[clip, trim=0 17 0 37pt, width=.7\textwidth]{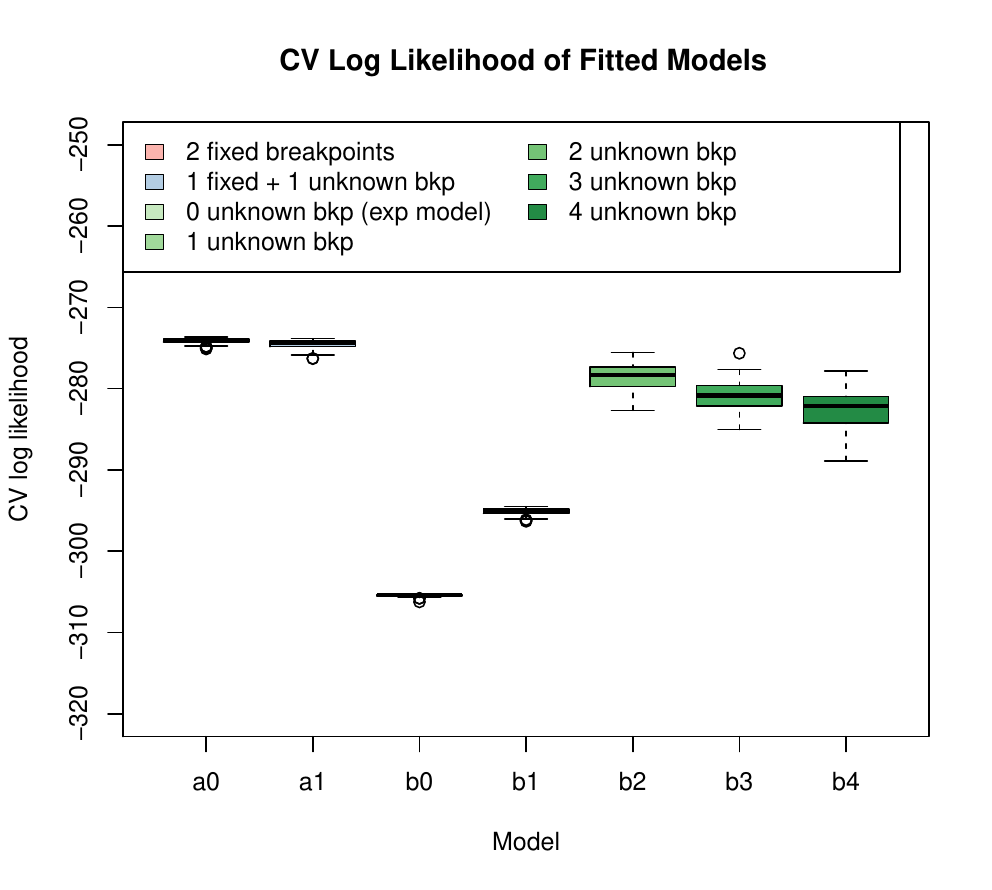}
  \caption{Cross validation log-likelihood of fitted PWE models}\label{fig:cv}
\end{figure}

The returned object \code{cv.pwexp.fit} is a vector of \code{nsim} CV log-likelihoods. We utilize a boxplot to visualize the CV log-likelihoods in Figure~\ref{fig:cv}.

\begin{verbatim}
R> boxplot(cv ~ model_ind, data.frame(cv = c(fit_a0_cv, fit_a1_cv, fit_b0_cv, fit_b1_cv, 
+                                            fit_b2_cv, fit_b3_cv, fit_b4_cv), 
+                                     model_ind = rep(model_ind, each = 100)), 
+          ylab = 'CV log likelihood', xlab = 'Model', ylim = c(-320,-250), 
+          col = col, main = 'CV Log Likelihood of Fitted Models')
R> legend('topleft', model_name, fill = col, ncol = 2, text.width = 3)
\end{verbatim}

Consistent with the BIC results, the model \code{fit\_a0} demonstrates the largest CV log-likelihood and serves as a baseline. Among the models without pre-specified change-points, \code{fit\_b2} continues to exhibit the largest CV log-likelihood.

Therefore,  we select model \code{fit\_b2} as our final optimal model for downstream analysis. The model details are stored in the object attributes list \code{attributes(fit\_b2)}. Specifically, we can extract the estimated hazard rates and change-points using \code{attr(fit\_b2, 'lam')} and \code{attr(fit\_b2, 'brk')}, respectively:
\begin{verbatim}
R> fit_b2

    brk1     brk2       lam1        lam2      lam3 likelihood      AIC      BIC
5.077201 13.95023 0.08955893 0.009611296 0.1907673  -1393.633 2797.266 2820.689

R> attr(fit_b2, 'lam')

[1] 0.089558930 0.009611296 0.190767295

R> attr(fit_b2, 'brk')

[1] 5.077201 13.950226
\end{verbatim}

\begin{figure}[!h]
\centering
  \includegraphics[clip, trim=0 16 0 37pt, width=.7\textwidth]{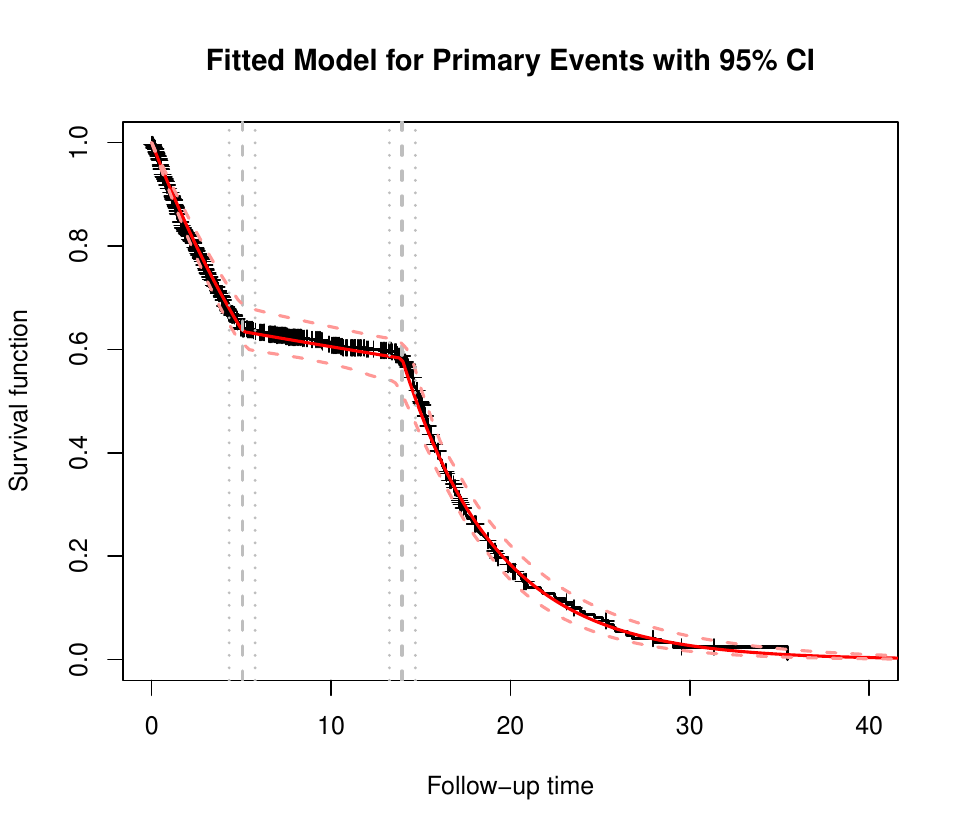}
  \caption{Visualization of the final model with $95\%$ CI}\label{fig:boot}
\end{figure}

We can utilize the \fct{boot.pwexp.fit} function to bootstrap this model and estimate the variance of the estimated parameters. The returned object includes a data frame  with \code{nsim} rows, each corresponding to estimated parameters in each bootstrapping sample. Similar to the \code{pwexp.fit} object, the bootstrapping model details are also restored in the object attributes list \code{attributes(fit\_b2\_boot)}. While bootstrapping the model is optional, it provides confidence intervals for event and timeline predictions, enhancing the reliability of our results. Figure~\ref{fig:boot} illustrates the fitted model with a $95\%$ confidence interval.

\begin{verbatim}
R> fit_b2_boot <- boot.pwexp.fit(fit_b2, nsim = 100, parallel = TRUE, mc.core = 10)
R> plot_survival(train$followT, train$event, xlim = c(0,40), 
+                main = 'Fitted Model for Primary Events with 95% CI')
R> plot_survival(fit_b2_boot, col = 'red', alpha = 0.05, CI_par = list(col = '#ff9896'))
R> brk_ci <- apply(attr(fit_b2_boot, 'brk'), 2, function(x)quantile(x, c(0.025, 0.975)))
R> abline(v = brk_ci, col = 'grey', lwd = 2, lty = 3)
\end{verbatim}

\subsection{Event/timeline Prediction}
As described in Section~\ref{sec:prediction}, for event/timeline prediction, a censoring model for drop-out and death (if present) can be estimated together with the event model. The censoring model is similar to the event model described in the previous section, but the response of the model is drop-out and death (if present) rather than event of interest. The censoring model is discretionary in scenarios where the drop-out (and death) rate is negligible. Nevertheless, in situations where the drop-out (and death) rates exhibit significance, constructing a censoring model becomes imperative  to ensure precise  event/timeline prediction. 

Here, we first create a variable \code{drop\_indicator} to be a time-to-event endpoint that represents the time of drop-out (and we do not have death in the simulated data), and
it will be used as the response variable in the censoring model. We fit an exponential censoring model with bootstrapping.

\begin{verbatim}
R> drop_indicator <- ifelse(train$censor_reason == 'drop_out' & 
+                             !is.na(train$censor_reason), 1, 0)
R> plot_survival(train$followT, drop_indicator, xlim = c(0, 40), 
+                ylab='Survival function of drop-out', 
+                main='Fitted Censoring Model for Drop-out with 95% CI')
R> fit_censor_boot <- boot.pwexp.fit(train$followT, drop_indicator, 
+                                    nbreak = 0, nsim = 100)
R> plot_survival(fit_censor_boot, col = 'red', alpha = 0.01, 
+                CI_par = list(col='#ff9896'))
\end{verbatim}

\begin{figure}[!h]
\centering
  \includegraphics[clip, trim=0 16 0 37pt, width=.7\textwidth]{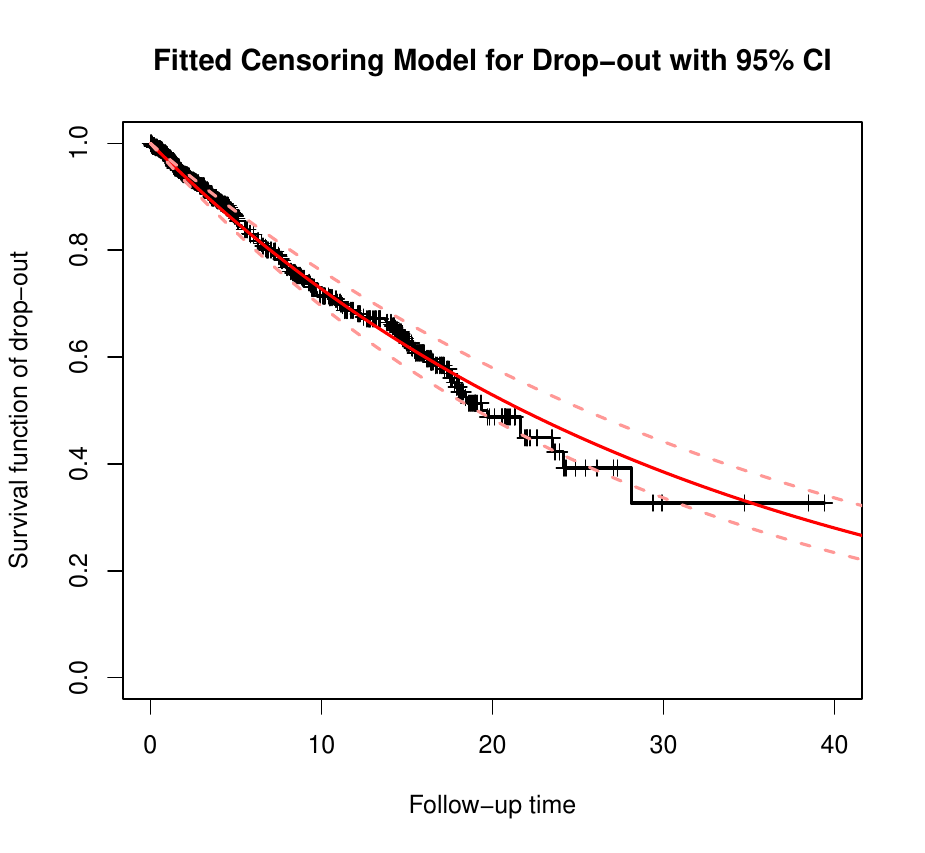}
  \caption{Visualization of the fitted censoring model for drop-out with $95\%$ CI}\label{fig:censor}
\end{figure}

Finally, we use the \fct{predict} function to predict the number of future events. The \code{future\_rand} argument defines the randomization curve of the remaining subjects. In this example, the future randomization rate remains $20$ subjects per month (\code{rand\_rate = 20}), and the total sample to be enrolled after the analysis time is the planned number of samples (\code{1000}) minus the number of randomized subjects at the analysis time (\code{NROW(train)}). The returned object from \fct{predict} contains predicted event curve functions, which can be used with the \fct{plot\_event} function to calculate the estimated events and plot the curve.

\begin{verbatim}
R> predicted_boot <- predict(fit_b2_boot, analysis_time = cut, 
+                            censor_model_boot = fit_censor_boot, 
+             future_rand = list(rand_rate = 20, total_sample = 1000 - NROW(train)), 
+                            n_each = 30)
\end{verbatim}

\begin{figure}[!t]
\centering
  \includegraphics[clip, trim=0 16 0 37pt, width=.7\textwidth]{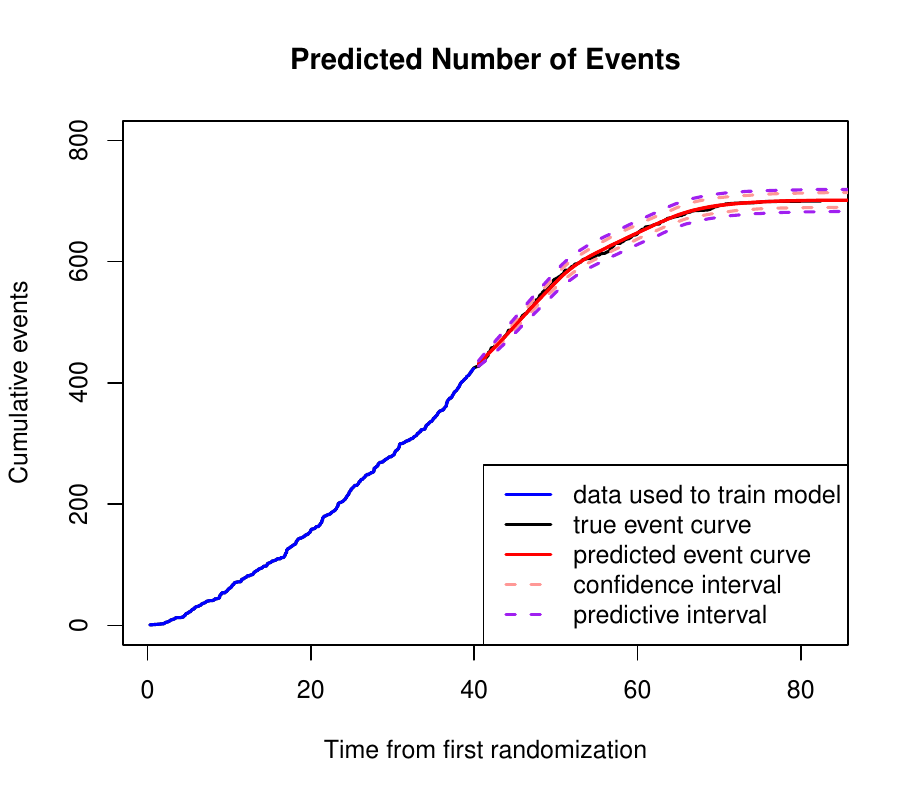}
  \caption{Predicted number of future events}\label{fig:pred1}
\end{figure}

We use \fct{plot\_event} function to plot the event curve with confidence interval (\code{type = 'confidence'}) and predictive interval (\code{type = 'predictive'}):
\begin{verbatim}
R> plot_event(dat$followT_abs, abs_time = T, event = dat$event, ylim = c(0, 800), 
+             main = 'Predicted Number of Events')
R> plot_event(train$followT_abs, abs_time = T, event = train$event, add = T, col='blue')
R> pred_event_confidence <- plot_event(predicted_boot, eval_at = seq(45, 90, 5), 
+                                      type = 'confidence')
R> pred_event_predictive <- plot_event(predicted_boot, eval_at = seq(45, 90, 5),
+                                    type = 'predictive', CI_par = list(col = 'purple'))
R> legend('bottomright', c('data used to train model', 'true event curve', 
+                          'predicted event curve', 'confidence interval', 
+                          'predictive interval'), lwd = 2, 
+               col=c('blue', 'black', 'red', '#ff9896', 'purple'), lty = c(1,1,1,2,2))
R> head(pred_event_confidence)

     time  n_event 5% n_event 95% n_event
[1,]   45 494.3127   489.2371    500.9097
[2,]   50 565.9164   557.6987    577.3319
[3,]   55 614.9518   604.7011    627.8907
[4,]   60 647.3800   637.2311    660.7467
[5,]   65 676.7368   666.0635    689.8273
[6,]   70 692.2490   680.6740    705.5199

R> head(pred_event_predictive)

     time  n_event 5% n_event 95% n_event
[1,]   45 494.4746   482.0519    507.2512
[2,]   50 566.0482   549.3727    584.1406
[3,]   55 614.8935   595.8857    634.8421
[4,]   60 647.5956   628.1870    666.9981
[5,]   65 677.5232   657.7572    696.6753
[6,]   70 693.0376   673.1747    711.9456
\end{verbatim}

The returned data frame contains the number of predicted events and its corresponding $95\%$ confidence/predictive interval at the specified times (by the \code{eval\_at} argument in the \fct{plot\_event} function). 

By setting argument \code{xyswitch = TRUE}, we can use the \fct{plot\_event} function to obtain the timeline for a given number of events in the future.  

\begin{figure}[!t]
\centering
  \includegraphics[clip, trim=0 16 0 37pt, width=.7\textwidth]{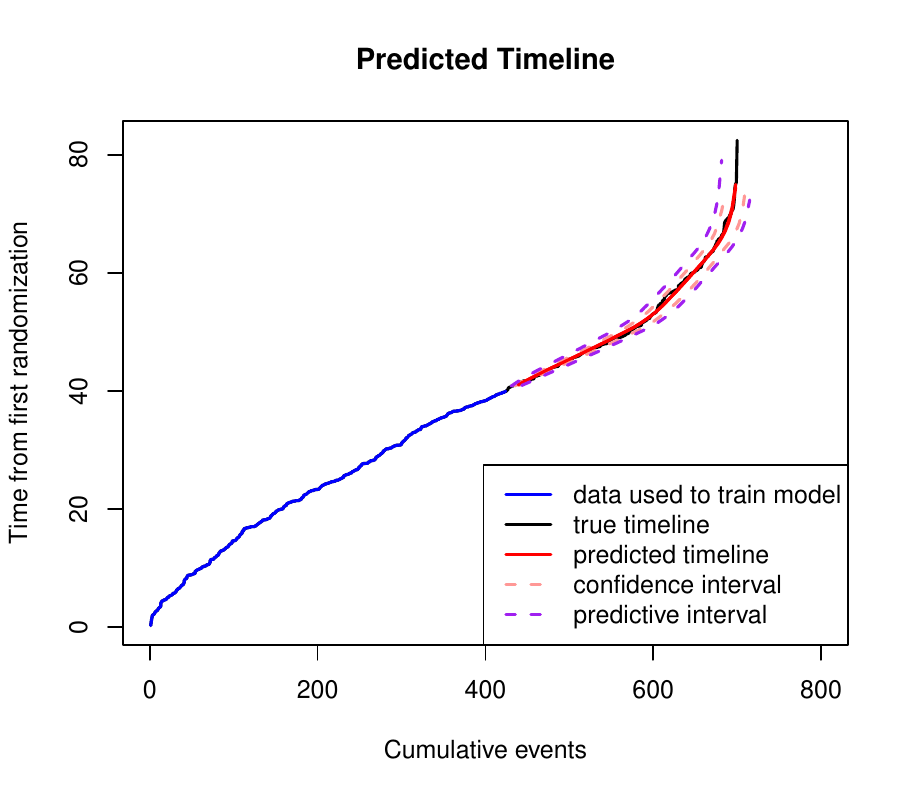}
  \caption{Predicted timeline for given number of future events}\label{fig:pred2}
\end{figure}

\begin{verbatim}
R> plot_event(dat$followT_abs, xlim = c(0, 800), event = dat$event, 
+             xyswitch = T, main = 'Predicted Timeline')
R> plot_event(train$followT_abs, abs_time = T, event=train$event, add = T, 
+             xyswitch = T, col = 'blue')
R> pred_time_confidence <- plot_event(predicted_boot, xyswitch = T, 
+                                     eval_at = seq(500, 700, 50), type = 'confidence')
R> pred_time_predictive <- plot_event(predicted_boot, xyswitch = T, 
+                                     eval_at = seq(500, 700, 50), type = 'predictive', 
+                                     CI_par = list(col = 'purple'))
R> legend('bottomright', c('data used to train model', 'true timeline', 
+                          'predicted timeline', 'confidence interval', 
+                          'predictive interval'), lwd = 2, 
+             col = c('blue', 'black', 'red', '#ff9896', 'purple'), lty = c(1,1,1,2,2))
R> head(pred_time_confidence)

     n_event     time  5% time 95% time
[1,]     500 45.37890 44.94330 45.79609
[2,]     550 48.85392 48.17232 49.38727
[3,]     600 53.05977 51.75140 54.30394
[4,]     650 60.41130 58.20409 62.13599
[5,]     700 77.63385 67.49335       NA

R> head(pred_time_predictive)

     n_event     time  5% time 95% time
[1,]     500 45.37776 44.53543 46.30582
[2,]     550 48.84987 47.72773 50.05325
[3,]     600 53.06970 51.16706 55.60011
[4,]     650 60.40710 57.27906 63.61008
[5,]     700 78.28629 65.73887       NA
\end{verbatim}

\section{Study Design Example}\label{sec:design}

In this section, we will demonstrate how to use the \pkg{PWEXP} package to assist in study design, such as sample size and study duration calculation, with one example. 

Assume we are planning a phase 3 trial design  with target population be patients newly diagnosed with diffuse large B-cell lymphoma and an international prognostic index (IPI) score of 4-5.  The primary objective of the trial is to demonstrate the superiority of the experimental treatment compared to the control arm in terms of overall survival (OS) endpoint. Additional characteristics for the study design are listed as below: 
\begin{itemize}
  \item subjects are randomized in a 1:1 ratio;
  \item a two-sided log-rank test will be employed, ensuring $90\%$ power at a significance level of $5\%$ (or one-sided at $2.5\%$);
  \item the target hazard ratio for the primary endpoint OS, is set to 0.6;
  \item a monthly dropout probability of 1\% is assumed;
  \item interim analyses are scheduled at $40\%$ and $70\%$ of total information;
  \item alpha-spending version of O'Brien-Fleming boundary for efficacy; no futility interim;
  \item a total of $660$ subjects will be recruited;
  \item recruitment begins at a rate 15 pt/month during the first 12 months; subsequently, increase of the number of sites and ramp up of recruitment by +6 pt/month each month until a maximum of 45 pt/month.
\end{itemize}

\subsection{Obtaining Model Assumption}
\begin{figure}[!h]
\centering
  \includegraphics[clip, trim=0 16 0 37pt, width=.7\textwidth]{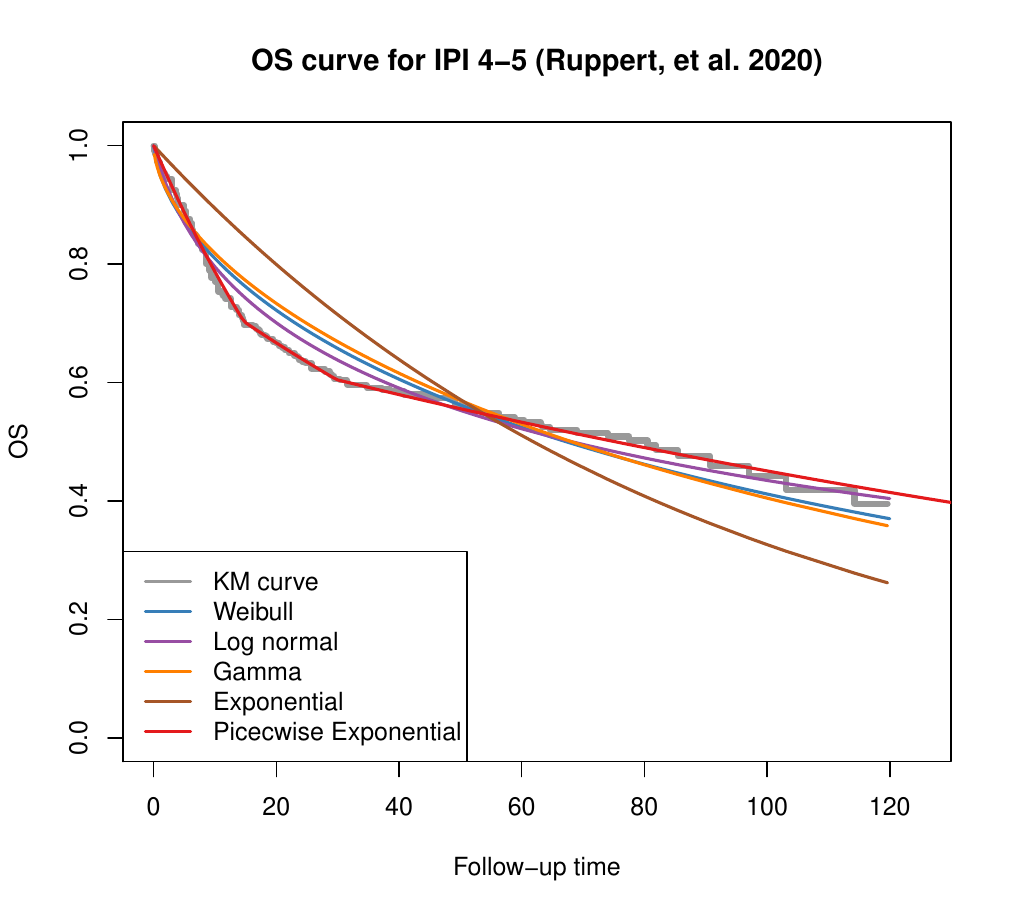}
  \caption{OS curve for IPI 4-5 subjects \citep{ruppert2020international}. Colored curves are fitted parametric models.}\label{fig:km}
\end{figure}

The initial step for study design involves deriving assumptions for the OS rate in the control arm by analyzing available historical data or literature such as  \citet{ruppert2020international}. To estimate the OS rate for the target population, we digitize the OS curves for IPI 4-5 (red curve) from Figure 2A using WebPlotDigitizer \citep{Rohatgi2024}. Subsequently, we reconstruct individual subject data (IPD) using the \pkg{IPDfromKM} package \citep{liu2021ipdfromkm}. The resulting OS curve for IPI 4-5 patients is represented by the gray curve in Figure~\ref{fig:km}. 

We explore various parametric models, including the exponential, Weibull, Gamma, and log-normal models, in an attempt to fit the data. However, these models prove inadequately capturing the KM curve. As a result, we consider employing a piecewise exponential model as a potential solution. 

\begin{verbatim}
R> dat <- read.csv(paste0("https://raw.githubusercontent.com/zjph602xtc/PWEXP/", 
+                  "main/docs/ruppert%20ipd.csv"))
R> PWE <- pwexp.fit(dat$Survival.time, dat$Status, nbreak = 2)
R> PWE

 brk1  brk2     lam1        lam2        lam3 likelihood      AIC      BIC
14.716 29.85 0.023956 0.009931584 0.004189957  -1065.844 2141.689 2161.984

R> PWEXP::ppwexp(12 * 1:4, rate = attr(PWE, 'lam'), breakpoint = attr(PWE, 'brk'), 
+         lower.tail = F)

[1] 0.7501575 0.6409900 0.5894241 0.5605208
\end{verbatim}

The red curve in Figure~\ref{fig:km} represents the fitted PWE model, which closely matches the actual KM curve. Based on the the fitted PWE model, the IPI 4-5 population group exhibits monthly hazard rates of $0.0240$, $0.00993$, and $0.00419$ for the time intervals of (0, $\leq$ 14.7 months), ($>$ 14.7 months, $\leq$ 29.9 months), and ($>$ 29.9 months), respectively. These hazard rates result in  a PFS rate of $75.0\%$, $64.1\%$, $58.9\%$, and $56.1\%$ at year 1, year 2, year 3, and year 4, respectively.

\subsection{Integration with Design Package}

Once OS rates were estimated by the fitted piecewise model \code{PWE}, they can be readily utilized in various clinical trial design packages. The model details are stored in the object attributes list \code{attributes(PWE)}. Specifically, we can extract the estimated hazard rates and change-points using \code{attr(PWE, 'lam')} and \code{attr(PWE, 'brk')}, respectively. As an illustration, we will demonstrate the integration with \pkg{rpact} package for study design:
\begin{verbatim}
R> design <- getDesignGroupSequential(
+    sided = 1, alpha = 0.025, beta = 0.1,
+    informationRates = c(0.4, 0.7, 1),
+    typeOfDesign = "asOF")
R> accrualTime <- getAccrualTime(accrualTime = c(0, 12:16), 
+                                accrualIntensity = c(15, 21, 27, 33, 39, 45),
+                                maxNumberOfSubjects = 660)
R> y <- getSampleSizeSurvival(
+    design = design, 
+    lambda2 = attr(PWE,'lam'),
+    dropoutRate1 = 0.01, dropoutRate2 = 0.01, dropoutTime = 1,
+    allocationRatioPlanned = 1,
+    accrualTime = accrualTime,
+    piecewiseSurvivalTime = c(0,attr(PWE,'brk')),
+    hazardRatio = 0.6)
R> summary(y)

Sample size calculation for a survival endpoint

Sequential analysis with a maximum of 3 looks (group sequential design), overall 
significance level 2.5% (one-sided).
The results were calculated for a two-sample logrank test, 
H0: hazard ratio = 1, 
H1: hazard ratio = 0.6, piecewise survival distribution, 
piecewise survival time = c(0, 14.716, 29.85), 
control lambda(2) = c(0.024, 0.01, 0.004), maximum number of subjects = 660, 
accrual time = c(12, 13, 14, 15, 16, 24), 
accrual intensity = c(15, 21, 27, 33, 39, 45), dropout rate(1) = 0.01, 
dropout rate(2) = 0.01, dropout time = 1, power 90%.

Stage                                         1      2      3 
Information rate                            40%    70%   100% 
Efficacy boundary (z-value scale)         3.357  2.445  2.001 
Overall power                            0.0981 0.6139 0.9000 
Number of subjects                        536.2  660.0  660.0 
Expected number of subjects under H1      647.8 
Cumulative number of events                65.3  114.3  163.4 
Analysis time                            21.248 27.089 35.146 
Expected study duration                    29.6 
Cumulative alpha spent                   0.0004 0.0074 0.0250 
One-sided local significance level       0.0004 0.0073 0.0227 
Efficacy boundary (t)                     0.436  0.633  0.731 
Exit probability for efficacy (under H0) 0.0004 0.0070 
Exit probability for efficacy (under H1) 0.0981 0.5157 

Legend:
  (t): treatment effect scale
\end{verbatim}
Based on the results, we will conduct two interim analyses at $21.2$ and $27.1$ months, with the final analysis scheduled for $35.1$ months to achieve an overall $90\%$ power. A total of $65.3$, $114.3$, $163.4$ cumulative events are expected to occur at each analysis time.

\subsection{Event/timeline and Follow-up Simulation}
The \pkg{PWEXP} package also offers simulation capabilities for study design purposes.
The \fct{sim\_followup} function can estimate the number of events at any specified time (\code{type = 'calendar'}). Additionally, it provides summary statistics of the follow-up time.  In the example below, the follow-up time is defined as the duration from randomization to censoring (either drop-out or end of the trial, whichever occurs first), specified by \code{follow\_up\_endpoint = c('drop\_out', 'cut')} \citep{betensky2015measures}. We can define summary statistics for the follow-up time. For instance, here we define \code{prop\_12} as the proportion of subjects whose follow-up time is larger than $12$ months. Moreover, we request statistics such as the \code{mean}, \code{median}, and \code{sum} of the follow-up time.

\begin{verbatim}
R> pfs_con <- function(n)PWEXP::rpwexp(n, rate = attr(PWE, 'lam'), 
+                                      breakpoint = attr(PWE, 'brk'))
R> pfs_trt <- function(n)PWEXP::rpwexp(n, rate = attr(PWE, 'lam') * 0.6, 
+                                      breakpoint = attr(PWE, 'brk'))
R> prop_12 <- function(x)mean(x >= 12)
R> sim_followup(at = c(21.248, 27.089, 35.146), type = 'calendar', 
+               group = c('trt', 'con'), allocation = 1, 
+               drop_rate = 0.01, by_group = T,
+               n_rand = c(rep(15, 12), 21, 27, 33, 39, rep(45, 8)), 
+               advanced_dist = list(event_dist = c(pfs_trt, pfs_con)),  
+               rep = 1000, stat = c(mean, median, sum, prop_12), 
+               follow_up_endpoint = c('cut', 'drop_out'))

$T_all
      at analysis_time   event subjects      mean    median       sum   prop_12
1 21.248        21.248  65.422   536.18  7.376394  5.662857  3954.951 0.2292295
2 27.089        27.089 114.560   660.00 11.076918  9.670736  7310.766 0.3550924
3 35.146        35.146 163.726   660.00 17.951402 17.106201 11847.925 0.8340242

$T_by_group
      at group analysis_time  event subjects      mean    median      sum   prop_12
1 21.248   con        21.248 40.165  268.285  7.379129  5.666560 1979.673 0.2293759
4 21.248   trt        21.248 25.257  267.895  7.373329  5.662474 1975.278 0.2290497
2 27.089   con        27.089 69.915  330.000 11.082933  9.680618 3657.368 0.3553515
5 27.089   trt        27.089 44.645  330.000 11.070902  9.661505 3653.398 0.3548333
3 35.146   con        35.146 99.260  330.000 17.949422 17.114265 5923.309 0.8343364
6 35.146   trt        35.146 64.466  330.000 17.953382 17.106924 5924.616 0.8337121
\end{verbatim}
The output displays the expected number of events at interim analyses (IAs) and final analysis (FA) as $65.4$, $114.6$, and $163.7$, respectively.  These figures closely align with the theoretical calculations performed using the \pkg{rpact} package. 

\begin{figure}[!h]
\centering
  \includegraphics[clip, trim=0 16 0 37pt, width=.7\textwidth]{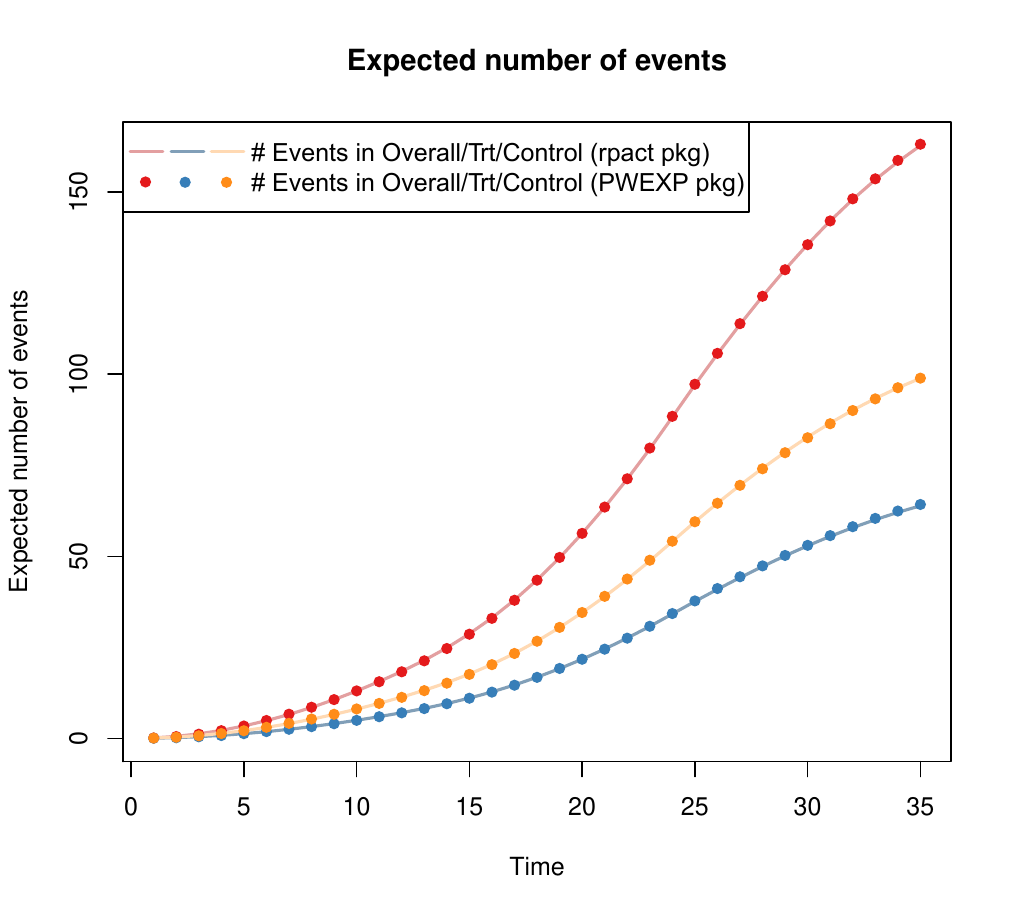}
  \caption{Expected number of events according to theoretical calculations from the \pkg{rpact} package and simulation results from the \pkg{PWEXP} package}\label{fig:events}
\end{figure}

If we plot the expected number of events according to theoretical calculations from the \pkg{rpact} package and simulation results from the \pkg{PWEXP} for each month in Figure~\ref{fig:events}, the simulation result is very accurate throughout the whole study duration. 
Furthermore, the simulation results provide summary statistics of follow-up for each arm at specified times. For example, at the time of the final analysis ($35.1$ months), $83.4\%$ of subjects will have a follow-up duration longer than $1$ year. This information is crucial for trial sponsors and health authorities in evaluating trial's conduct and performance.

The \fct{sim\_followup} function can also estimate the timeline based on a specified number of randomized subjects or number of events (\code{type = 'sample'} or \code{type = 'event'}). The following example demonstrates the analysis time for every $10\%$ increase in information fraction from $10\%$ to $100\%$:
\begin{verbatim}
R> sim_followup(at = seq(0.1, 1, 0.1) * 163.4, type = 'event', group = c('trt', 'con'),
+               allocation = 1, drop_rate = 0.01,
+               n_rand = c(rep(15, 12), 21, 27, 33, 39, rep(45, 8)), 
+               advanced_dist = list(event_dist = c(pfs_trt, pfs_con)), rep = 1000)

$T_all
    at analysis_time event subjects      mean    median        sum
1   16      11.03759    16  166.952  5.280129  5.174119   894.3154
2   33      15.90078    33  299.653  6.350680  5.413976  1906.5474
3   49      18.83120    49  427.435  6.744884  5.033877  2891.7742
4   65      21.16049    65  532.298  7.359293  5.650851  3929.0303
5   82      23.23549    82  620.501  8.093451  6.535725  5034.8399
6   98      25.09083    98  656.938  9.317254  7.851815  6124.7430
7  114      27.04290   114  659.976 11.028731  9.625848  7278.7666
8  131      29.35984   131  660.000 13.060754 11.770090  8620.0980
9  147      31.95780   147  660.000 15.279185 14.168758 10084.2622
10 163      35.37393   163  660.000 18.095337 17.312156 11942.9226
\end{verbatim}

\section{Summary and Discussion} \label{sec:summary}
Survival analysis plays a crucial role in various fields, where understanding the time until an event of interest occurs is the primary goal. The \pkg{PWEXP} package presented in this paper offers a comprehensive suite of tools tailored specifically for survival data with a piecewise exponential distribution (PWE). 
The \pkg{PWEXP} package follows a modular and user-friendly design, making it accessible to both novice and experienced researchers. 
The package is able to automatically fit the change-points, assess model performance using statistical criteria like AIC, BIC, and cross-validation log-likelihood, and estimate parameter uncertainty through bootstrapping. In addition to model estimation, the package supports event/timeline simulation and prediction. These tools empower researchers to explore complex survival datasets, derive meaningful insights, and make informed decisions. Additionally, the package's integration with other statistical tools, such as \pkg{rpact}, \pkg{gsDesign}, enables researchers to perform comprehensive trial design simulations, including power calculations, interim analyses, and timeline estimations.

There are several future development directions worth studying. For example, current event/timeline prediction methods assume timely and complete reporting of data,  which may not always align with real-world scenarios. Incorporating event report lag considerations, as explored in previous studies \citep{wang2012predicting}, could enhance prediction accuracy. 
Additionally, accounting for factors like cross-over (e.g., switch from control to treatment arm) and treatment switches (e.g., receive subsequent therapy) would broaden the package's applicability in diverse research contexts.


-------------------------------------

\section*{Computational details}

The results in this paper were obtained using \proglang{R}~4.3.0 with the \pkg{PWEXP}~0.5.0 package. \proglang{R} itself and all packages used are available from the Comprehensive
\proglang{R} Archive Network (CRAN) at \url{https://CRAN.R-project.org/}.


\bibliographystyle{Chicago}
\bibliography{article}

@Manual{pch,
    title = {\pkg{pch}: Piecewise Constant Hazard Models for Censored and Truncated Data},
    author = {Paolo Frumento},
    year = {2021},
    note = {\proglang{R}~package~version~2.0},
    url = {https://CRAN.R-project.org/package=pch},
}

@Manual{muhaz,
    title = {\pkg{muhaz}: Hazard Function Estimation in Survival Analysis},
    author = {Kenneth Hess and Gentleman Robert},
    year = {2021},
    note = {\proglang{R}~package~version~1.2.6.4},
    url = {https://CRAN.R-project.org/package=muhaz},
}

@Manual{eventTrack,
    title = {\pkg{eventTrack}: Event Prediction for Time-to-Event Endpoints},
    author = {Kaspar Rufibach},
    year = {2022},
    note = {\proglang{R}~package~version~1.0.2},
    url = {https://CRAN.R-project.org/package=eventTrack},
}

@Article{PiecewiseChangepoint,
    title = {Change-Point Detection for Piecewise Exponential Models},
    author = {Philip Cooney and Arthur White},
    journal = {arXiv},
    year = {2021},
    pages = {1--21},
    url = {https://arxiv.org/abs/2112.03962},
}

@Manual{rpact,
    title = {\pkg{rpact}: Confirmatory Adaptive Clinical Trial Design and Analysis},
    author = {Gernot Wassmer and Friedrich Pahlke},
    year = {2024},
    note = {\proglang{R}~package~version~3.5.1},
    url = {https://CRAN.R-project.org/package=rpact},
}

@Manual{geDesign,
    title = {\pkg{gsDesign}: Group Sequential Design},
    author = {Keaven Anderson},
    year = {2023},
    note = {\proglang{R}~package~version~3.6.0},
    url = {https://CRAN.R-project.org/package=gsDesign},
}

@misc{Rohatgi2024,
  url = {https://automeris.io/WebPlotDigitizer},
  author = {Rohatgi, Ankit},
  title = {Webplotdigitizer: Version 4.7},
  year = {2024}
}

@book{loubert1986inference,
  title={Inference Procedures for the Piecewise Exponential Model When the Data Are Arbitrarily Censored},
  author={Loubert, Sharon K},
  year={1986},
  publisher={Iowa State University}
}

@article{betensky2015measures,
  title={Measures of Follow-pp in Time-to-event Studies: Why Provide Them and What Should They Be?},
  author={Betensky, Rebecca A},
  journal={Clinical Trials},
  volume={12},
  number={4},
  pages={403--408},
  year={2015},
  publisher={SAGE Publications Sage UK: London, England}
}

@misc{kuchenhoff1996exact,
          volume = {27},
           title = {An Exact Algorithm for Estimating Breakpoints in Segmented Generalized Linear Models},
          author = {Helmut K\"uchenhoff},
          series = {sfb386},
            year = {1996},
             url = {http://nbn-resolving.de/urn/resolver.pl?urn=nbn:de:bvb:19-epub-1429-2}
}

@article{wang2012predicting,
  title={Predicting Analysis Time in Event-Driven Clinical Trials with Event-Reporting Lag},
  author={Wang, Jianming and Ke, Chunlei and Jiang, Qi and Zhang, Charlie and Snapinn, Steven},
  journal={Statistics in Medicine},
  volume={31},
  number={9},
  pages={801--811},
  year={2012},
  publisher={Wiley Online Library}
}

@article{liu2021ipdfromkm,
  title={IPDfromKM: Reconstruct Individual Patient Data from Published Kaplan-Meier Survival Curves},
  author={Liu, Na and Zhou, Yanhong and Lee, J Jack},
  journal={BMC medical Research Methodology},
  volume={21},
  number={1},
  pages={111},
  year={2021},
  publisher={Springer}
}

@article{muggeo2003estimating,
  title={Estimating Regression Models with Unknown Break-Points},
  author={Muggeo, Vito MR},
  journal={Statistics in Medicine},
  volume={22},
  number={19},
  pages={3055--3071},
  year={2003},
  publisher={Wiley Online Library}
}

@article{ruppert2020international,
  title={International Prognostic Indices in Diffuse Large B-cell Lymphoma: A Comparison of IPI, R-IPI, and NCCN-IPI},
  author={Ruppert, Amy S and Dixon, Jesse G and Salles, Gilles and Wall, Anna and Cunningham, David and Poeschel, Viola and Haioun, Corinne and Tilly, Herve and Ghesquieres, Herve and Ziepert, Marita and others},
  journal={Blood, The Journal of the American Society of Hematology},
  volume={135},
  number={23},
  pages={2041--2048},
  year={2020},
  publisher={American Society of Hematology Washington, DC}
}

@article{bagiella2001predicting,
  title={Predicting Analysis Times in Randomized Clinical Trials},
  author={Bagiella, Emilia and Heitjan, Daniel F},
  journal={Statistics in medicine},
  volume={20},
  number={14},
  pages={2055--2063},
  year={2001},
  publisher={Wiley Online Library}
}

@article{ying2008weibull,
  title={Weibull Prediction of Event Times in Clinical Trials},
  author={Ying, Gui-shuang and Heitjan, Daniel F},
  journal={Pharmaceutical Statistics: The Journal of Applied Statistics in the Pharmaceutical Industry},
  volume={7},
  number={2},
  pages={107--120},
  year={2008},
  publisher={Wiley Online Library}
}

@book{klein2003survival,
  title={Survival Analysis: Techniques for Censored and Truncated Data},
  author={Klein, John P and Moeschberger, Melvin L and others},
  volume={1230},
  year={2003},
  publisher={Springer}
}

@incollection{qian2013multiple,
  title={Multiple Change-Point Detection in Piecewise Exponential Hazard Regression Models with Long-Term Survivors and Right Censoring},
  author={Qian, Lianfen and Zhang, Wei},
  booktitle={Contemporary Developments in Statistical Theory: A Festschrift for Hira Lal Koul},
  pages={289--304},
  year={2013},
  publisher={Springer}
}

@article{matthews1982testing,
  title={On Testing for a Constant Hazard against a Change-Point Alternative},
  author={Matthews, David E and Farewell, Vernon T},
  journal={Biometrics},
  pages={463--468},
  year={1982},
  publisher={JSTOR}
}

@article{henderson1990problem,
  title={A Problem with the Likelihood Ratio Test for a Change-Point Hazard Rate Model},
  author={Henderson, Robin},
  journal={Biometrika},
  volume={77},
  number={4},
  pages={835--843},
  year={1990},
  publisher={Oxford University Press}
}

@article{goodman2011detecting,
  title={Detecting Multiple Change Points in Piecewise Constant Hazard Functions},
  author={Goodman, Melody S and Li, Yi and Tiwari, Ram C},
  journal={Journal of Applied Statistics},
  volume={38},
  number={11},
  pages={2523--2532},
  year={2011},
  publisher={Taylor \& Francis}
}

@article{dupuy2006estimation,
  title={Estimation in a Change-Point Hazard Regression Model},
  author={Dupuy, Jean-Fran{\c{c}}ois},
  journal={Statistics \& Probability Letters},
  volume={76},
  number={2},
  pages={182--190},
  year={2006},
  publisher={Elsevier}
}

@article{yao1986maximum,
  title={Maximum Likelihood Estimation in Hazard Rate Models with a Change-Point},
  author={Yao, Yi-Ching},
  journal={Communications in Statistics-Theory and Methods},
  volume={15},
  number={8},
  pages={2455--2466},
  year={1986},
  publisher={Taylor \& Francis}
}

@article{kim2020bayesian,
  title={Bayesian Multiple Change-Points Estimation for Hazard with Censored Survival Data from Exponential Distributions},
  author={Kim, Jaehee and Cheon, Sooyoung and Jin, Zhezhen},
  journal={Journal of the Korean Statistical Society},
  volume={49},
  pages={15--31},
  year={2020},
  publisher={Springer}
}

@article{chapple2020novel,
  title={A Novel Bayesian Continuous Piecewise Linear Log-Hazard Model, with Estimation and Inference via Reversible Jump Markov Chain Monte Carlo},
  author={Chapple, Andrew G and Peak, Taylor and Hemal, Ashok},
  journal={Statistics in Medicine},
  volume={39},
  number={12},
  pages={1766--1780},
  year={2020},
  publisher={Wiley Online Library}
}

@article{cooney2021change,
  title={Change-Point Detection for Piecewise Exponential Models},
  author={Cooney, Philip and White, Arthur},
  journal={arXiv preprint arXiv:2112.03962},
  year={2021}
}

@article{li2013estimation,
  title={Estimation in a Change-Point Hazard Regression Model with Long-Term Survivors},
  author={Li, Yunxia and Qian, Lianfen and Zhang, Wei},
  journal={Statistics \& Probability Letters},
  volume={83},
  number={7},
  pages={1683--1691},
  year={2013},
  publisher={Elsevier}
}

@article{friedman1982piecewise,
  title={Piecewise Exponential Models for Survival Data with Covariates},
  author={Friedman, Michael},
  journal={The Annals of Statistics},
  volume={10},
  number={1},
  pages={101--113},
  year={1982},
  publisher={Institute of Mathematical Statistics}
}

@Manual{R,
  title = {\proglang{R}: {A} Language and Environment for Statistical Computing},
  author = {{\proglang{R} Core Team}},
  organization = {\proglang{R} Foundation for Statistical Computing},
  address = {Vienna, Austria},
  year = {2017},
  url = {https://www.R-project.org/},
}


\newpage

\begin{appendix}

\section[Generating Synthetic Survival Data with simdata()]{Generating Synthetic Survival Data with \fct{simdata}} \label{app:simudata}

\begin{figure}[!ht]
    \centering
    \includegraphics[width=\textwidth]{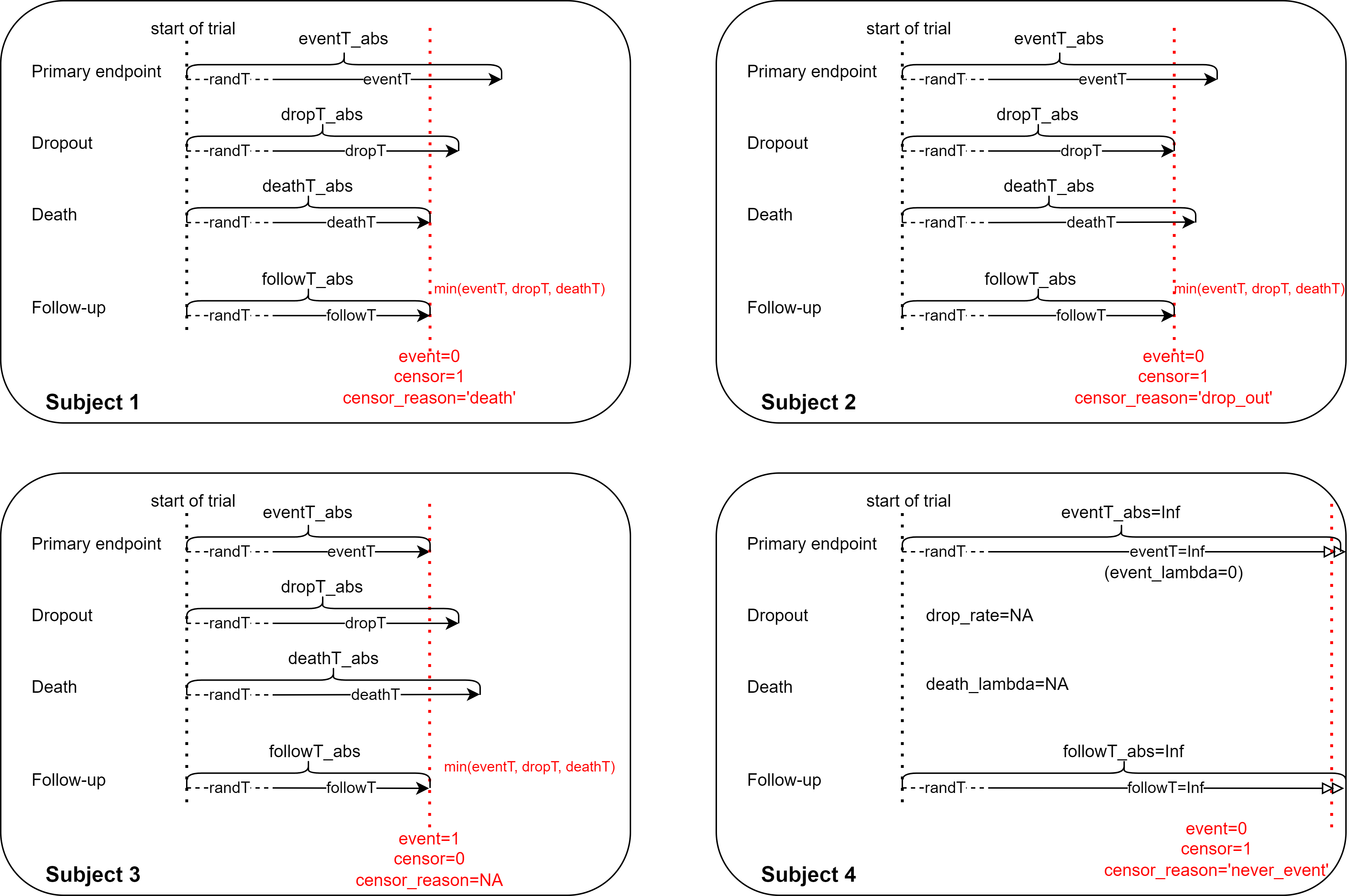}
    \caption{Relationship between variables in the generated dataset}
    \label{fig:simudata}
\end{figure}

The \fct{simdata} function can generate very flexible synthetic survival dataset with:
\begin{itemize}
  \item randomization/enrollment time defined by randomization/enrollment rate (number of subjects per month) with \code{rand\_rate} argument, or randomization/enrollment curve (number of subjects in each month) wiht \code{n\_rand} argument;
  \item multiple treatment groups with stratification by user-defined allocation ratio with \code{group}, \code{strata} and \code{allocation} arguments;
  \item primary endpoint (event), drop-out and death with exponential distribution with \code{event\_lambda}, \code{drop\_rate}, \code{death\_lambda} arguments, or user-defined distributions (e.g., piecewise exponential, mixture distribution, etc.) with \code{advanced\_dist} argument for each stratification in each group.
\end{itemize}
            
In the generated dataset, all subjects are followed for infinite time and we can truncate it by a clinical cut-off date (DCOD) with the \fct{cut\_dat} function later. The following variables are included the generated dataset (some are requested by the \code{add\_column} argument):
\begin{itemize}
  \item \code{randT} is the randomization time for each subject;
  \item \code{eventT}, \code{dropT}, \code{deathT} are the event time, drop-out time and death time, respectively;
  \item \code{followT} is the follow-up time, which is the minimum value of \code{eventT}, \code{dropT} and \code{deathT};
  \item \code{eventT\_abs}, \code{dropT\_abs}, \code{deathT\_abs} and \code{followT\_abs} are the corresponding time from the beginning of the trial (e.g., \code{eventT\_abs} is the sum of \code{randT} and \code{eventT}, etc.);
  \item \code{event} indicates whether the primary event occurred at the end of follow-up with value $1$ for occurrence of event. \code{censor} indicates whether the censoring (including drop-out and death) occurred at the end of follow-up with value $1$ for censoring. If a subject is censored, \code{censor\_reason} shows the type of censoring (i.e., \code{drop\_out}, \code{death} or \code{never\_event} (when \code{followT == Inf})).
\end{itemize}
Figure~\ref{fig:simudata} shows some examples of the relationship between these variables.

\end{appendix}


\end{document}